\documentclass[twocolumn,showpacs,aps,prd,nobibnotes,nofootinbib,floatfix]{revtex4-2}

\usepackage{amsmath}
\usepackage{graphicx,subfigure}
\usepackage[usenames]{color}
\usepackage[normalem]{ulem}
\usepackage{textcomp}
\usepackage{gensymb}
\usepackage{xspace}
\usepackage{verbatim}

\usepackage{soul}

\usepackage[colorlinks=true]{hyperref}
\hypersetup{
  citecolor  =red,
  urlcolor=blue
}
\usepackage{orcidlink}
\newcommand{\rh}{r_{\text{h}}}
\newcommand{\dd}{\text{d}}
\newcommand*\abs[1]{\lvert#1\rvert}
\begin{document}
%================================================

%================================================
\title{Test fields and naked singularities: is the second law the cosmic censor?}
%================================================

%================================================
\author{Si-Jiang Yang\orcidlink{0000-0002-8179-9365}$^a$$^b$}
\email{yangsj@lzu.edu.cn}
\author{Zheng-Quan Cui\orcidlink{0000-0003-2140-5077}$^c$}
\email{cuizhq5@mail.sysu.edu.cn}
\author{Shao-Wen Wei\orcidlink{0000-0003-0731-0610}$^a$$^b$}
\email{weishw@lzu.edu.cn}
\author{Yu-Xiao Liu\orcidlink{0000-0002-4117-4176}$^a$$^b$}%
\email{liuyx@lzu.edu.cn, corresponding author}
%================================================
\affiliation{$^{a}$Lanzhou Center for Theoretical Physics, Key Laboratory of Theoretical Physics of Gansu Province, Key Laboratory of Quantum Theory and Applications of MoE, Gansu Provincial Research Center for Basic Disciplines of Quantum Physics, Lanzhou University, Lanzhou 730000, China\\
$^{b}$Institute of Theoretical Physics $\&$ Research Center of Gravitation, School of Physical Science and Technology, Lanzhou University, Lanzhou 730000, China\\
$^{c}$School of Physics and Astronomy, Sun Yat-Sen University,
Zhuhai 519082, China}
%================================================
\date{\today}
%================================================

%================================================
\begin{abstract}
It has been claimed that a Kerr-Newman black hole can generically be overspun by neutral test fields, and it has been argued that even when backreactions are taken into account, the black hole can still be destroyed. In this paper, we revisit the weak cosmic censorship conjecture for a Kerr-Newman black hole with a test scalar field and a neutrino field, and point out that the assumption in previous work regarding the energy and angular momentum of the test fields absorbed by the black hole violates the second law of black hole thermodynamics. By solving the test scalar field and neutrino field near the event horizon explicitly, we demonstrate that an extremal Kerr-Newman black hole cannot be overspun by a test scalar field but can be destroyed by a neutrino field. Our results indicate that the condition required to overspin an extremal Kerr-Newman black hole coincides with the condition needed to violate the second law of black hole thermodynamics. Furthermore, we observe that such a violation of the second law might inevitably result in a breakdown of the weak cosmic censorship conjecture.
\end{abstract}
%================================================
\maketitle
%================================================

%================================================
\section{Introduction}\label{sec:intro}
%================================================

Black holes are the simplest and most interesting objects predicted by general relativity. The most general stationary electrovacuum solution to Einstein's equations in a four-dimensional asymptotically flat spacetime is the Kerr-Newman geometry, which is uniquely characterized by its mass $M$, angular momentum $J$, and charge $Q$~\cite{Ruffini:1971bza}. Though black holes have curvature singularities, the weak cosmic censorship conjecture states that these singularities should be hidden behind the event horizon and can never be seen by distant observers~\cite{Penr69}. Another intriguing feature is that black holes have four laws of thermodynamics~\cite{Bardeen:1973gs}. The second law of black hole thermodynamics states that the entropy is proportional to the area of the event horizon and can never decrease~\cite{Bekenstein:1973ur}.  
Naively, one expects there are relationships between the weak cosmic censorship conjecture and the second law of black hole thermodynamics.

The weak cosmic censorship conjecture has been examined through various approaches, such as numerical evolution of collapsing matter fields~\cite{Chri84,OrPi87,Lemo92}, numerical simulations of perturbed
black holes~\cite{Corelli:2021ikv,Eperon:2019viw,HSSLW20,Song:2022dpw,Cui:2024xqg}, and numerical evolution of
collision and merge of two black holes~\cite{SCPBHY09,AELL19,AELL19a}. One particularly intriguing method of testing the conjecture involves the gedanken experiment aimed at destroying the event horizon. A typical strategy of the gedanken experiment is to throw test particles or fields into an extremal or near-extremal black hole and check whether the event horizon exists or not for the final composite object.  While numerous examples have been explored to assess the validity of the weak cosmic censorship conjecture, the precise conditions under which it holds continue to be an area of active research. 

The pioneering work of the gedanken experiment to test the weak cosmic censorship conjecture was proposed by Wald~\cite{Wald74}. By throwing a test particle with large charge or angular momentum into an extremal Kerr-Newman black hole, Wald found that particles that cause the destruction of the event horizon can never be absorbed by the black hole. 
However, further studies showed that near-extremal black holes can ``jump over" the extremal limit, becoming naked singularities by absorbing test particles with large charge or angular momentum~\cite{Hube99,Hod02}. Hubeny showed that a near-extremal Reissner-Nordstr\"om black hole can be destroyed by a charged test particle~\cite{Hube99}. Similar work found that the event horizon of a near-extremal Kerr black hole can also be overspun~\cite{JaSo09}. Other black holes in modified gravity or gravity theories with nonlinear electromagnetic field also gave the same results~\cite{Yang:2022yvq,Ghosh:2019dzq,Liu:2020cji,Miao:2024wwd}. Further investigation suggested that when higher orders for the energy, angular momentum and charge of the test particles are taken into account, even extremal black holes might still be destroyed~\cite{Gao:2012ca,Yang:2020czk}. But when backreaction and finite-size effects are taken into account, those counterexamples appeared to be rescued~\cite{BaCK10,BaCK11,ZVPH13,CoBa15,Gwak17,LiWL19}. Recently, Sorce and Wald proposed a new kind of gedanken experiment by taking into account the second-order perturbations; they found that matter fields satisfying the null energy condition cannot destroy the Kerr-Newman black hole~\cite{SoWa17}. Subsequent systematic studies further support the result that the event horizon cannot be destroyed by this kind of gedanken experiment~\cite{SaJi21,ZhJi20,QYWR20,WaJi20,JiGa20,ChLN19,Shaymatov:2020byu}.

Recently, it has been shown that the Kerr-Newman black hole can be generically overspun by a classical test scalar field and a neutrino field~\cite{Duztas:2019ick}. The author claimed that the destruction of the event horizon for neutrino fields cannot be fixed by any form of backreaction effects, and therefore concluded that such a process may serve as a counterexample to the weak cosmic censorship conjecture. It is a manifest contradiction with other works~\cite{Gwak:2021tcl,NaQV16} that testing classical fields cannot destroy Kerr-Newman black holes. One of our goal is thus to provide possible solutions to the contradiction by arguing that the assumption of the previous work~\cite{Duztas:2019ick} for the energy and angular momentum absorbed by the black hole violates Hawking's area-increasing theorem and solves the problem why the work leads to the conclusion that the weak cosmic censorship conjecture can be violated for the Kerr-Newman black hole. Black hole thermodynamics and the weak cosmic censorship conjecture are both fundamental to black hole physics, and they have a close relationship. Research has investigated the validity of the weak cosmic censorship conjecture and discussed its relationship with the first and third laws of black hole thermodynamics~\cite{Yoshida:2024txh,Yang:2023hll,Yang:2020iat,Feng:2020tyc,Gao:2022ckf}, and the relationship between the second law and the weak cosmic censorship conjecture is still an active area of research~\cite{Lin:2024deg,Ali:2023iuz}. One of our other goals is thus to analyze the relationship between the second law of black hole thermodynamics and the weak cosmic censorship conjecture, and argue that the violation of the second law of black hole thermodynamics might inevitably lead to the violation of the weak cosmic censorship conjecture.

The outline of the paper is as follows. In Sec.\,\ref{sec:2}, we analysis the previous work to show that the key assumption for the energy and angular momentum absorbed by the black hole violates the second law of black hole thermodynamics. In Sec.\,\ref{sec:3} we check the validity of the weak cosmic censorship conjecture for extremal Kerr-Newman black holes with a test scalar field and a test neutrino field. In Sec.\,\ref{Sec:general}, we analyze the relationship between the second law of black hole thermodynamics and the weak cosmic censorship conjecture. The last section is devoted to the conclusion and discussion.
%=====================================================

%=====================================================
\section{Thermodynamics and the weak cosmic censorship conjecture}\label{sec:2}
%=====================================================

%=====================================================
\subsection{The Kerr-Newman black hole and its thermodynamics}
%=====================================================

The four-dimensional Kerr-Newman spacetime is an electrovacuum solution of Einstein's field equations.
The metric for the Kerr-Newman spacetime in Boyer-Lindquist coordinates can be written in the form
\begin{equation}
  \begin{split}
     ds^2 &=\frac{\Delta}{\Sigma}\left(\mathrm{d}t-a\sin^2\theta \,\mathrm{d}\phi\right)^2 -
  \frac{\Sigma}{\Delta} \,\mathrm{d}r^2  \\
  &-\Sigma \,\mathrm{d}\theta^2
  -\frac{\sin^2\theta}{\Sigma}\left[a \,\mathrm{d}t
  -\left(r^2+a^2\right) \,\mathrm{d}\phi\right]^2,
  \end{split}
\end{equation}
and the electromagnetic field potential is given by
\begin{equation}\label{Apotential}
  A=-\frac{Qr(\dd t-a\sin^2\theta \,\dd\phi )}{\Sigma},
\end{equation}
where the metric functions are given by
\begin{align}\label{Mfunction}
  \Delta & =r^2-2Mr+a^2+Q^2, \\
  \Sigma & =r^2+a^2\cos^2\theta,
\end{align}
and $a=J/M$ is the angular momentum per unit mass, $M$ and $Q$ are the mass and electric charge of the spacetime, respectively. The event horizon of the black hole is $
\rh=M+\sqrt{M^2-a^2-Q^2}$. Whether or not there is an event horizon is determined by the minimum of the metric function
\begin{equation}
    \Delta_{\text{min}}=-M^2+a^2+Q^2.\label{DeltaminH}
\end{equation}
Only when the minimum of the metric function is nonpositive does the metric describe a Kerr-Newman black hole; otherwise it describes a naked singularity.

The thermodynamic quantities of the black hole are
\begin{align}
    T_{\text{H}}&=\frac{1}{4\pi \rh}\frac{\rh^2-a^2-Q^2}{\rh^2+a^2}, \quad S_{\text{BH}}=\pi(\rh^2+a^2),\\
    \phi_{\text{H}}&=\frac{Q\rh}{\rh^2+a^2},  \qquad   \qquad \qquad    \Omega_{\text{H}}=\frac{a}{\rh^2+a^2}, 
\end{align}
where $T_{\text{H}}$, $S_{\text{BH}}$, $\phi_{\text{H}} $, and $\Omega_{\text{H}}$ are the Hawking temperature, entropy, electro-potential and angular velocity of the black hole, respectively.
The first law and the Smarr relation of the black hole are
\begin{align}
dM&=T_{\text{H}}dS_{\text{BH}}+\phi_{\text{H}}dQ+\Omega_{\text{H}}dJ,\\
M&=2T_{\text{H}}S_{\text{BH}}+\phi_{\text{H}}Q+2\Omega_{\text{H}}J.
\end{align}

%====================================
\subsection{Previous work: violate the second law of black hole thermodynamics}\label{sec:2b}
%====================================

In a recent work~\cite{Duztas:2019ick}, the author claimed that the Kerr-Newman black hole can be generically overspun by a test scalar and a test neutrino field, and argued that the destruction of the event horizon for a neutrino field cannot be fixed by any form of backreaction effects due to the absence of superradiace. The author therefore concluded that such a process may serve as a counterexample to the weak cosmic censorship conjecture~\cite{Duztas:2019ick}. The analysis of the work is based on the assumption that the energy of the wave packet absorbed by the black hole is
\begin{equation}
\delta E=M_0\epsilon,
\end{equation}
where $M_0$ is the initial mass of the black hole and $\epsilon$ is an infinitesimal constant.  The ratio of the angular momentum $\delta J$ to the energy $\delta E$ absorbed by the black hole, as Bekenstein argued from the fluxes of the quantized wave at infinity \cite{Bekenstein:1973mi}, must be the ratio of the azimuthal number $m$ to the wave frequency $\omega$. Thus the angular momentum carried by the field across the horizon is
\begin{equation}
\delta J =\frac{m}{\omega}\delta E.
\end{equation}
Plugging these into the black hole horizon checking condition~\eqref{DeltaminH}, indeed, we can get the result that there exists an allowed range of frequencies to overspin a Kerr-Newman black hole.

Whether or not we can use the laws of black hole thermodynamics to investigate the weak cosmic censorship conjecture for extremal and near-extremal black holes is a controversial issue, since the final composite body might not be a black hole.  Nat\'ario et al. relied heavily on black hole thermodynamics to prove that extremal Kerr-Newman or Kerr-Newman anti-de Sitter black holes cannot be destroyed by test fields~\cite{NaQV16}. Gwak's early work \cite{Gwak17} and the work of Liang et al. \cite{LiWL19} used the increase of black hole entropy to derive the low bound of the energy for the particle to destroy near extremal Kerr-Sen and Kerr-MOG black holes. Lin et al. used the second law of black hole thermodynamics to demonstrate the validation of the weak cosmic censorship conjecture in higher derivative gravity theory~\cite{Lin:2022ndf}. However, recent work of Gwak argued that one must not impose the laws of thermodynamics to investigate the weak cosmic censorship conjecture in the extremal and near extremal black holes \cite{Gwak18}. To avoid this dilemma, we proceed our argument by assuming that the initial black hole configuration is far from extremal. 

As argued in the work~\cite{Duztas:2019ick}, in the absence of supper-radiance for neutrino field scattering, a Kerr-Newman black hole can generically be overspun provided the frequency of the incoming scalar or neutrino fields $\omega < \omega_c$ and the energy $ \delta E \sim M_0 \epsilon $ , where $\omega_c$ is the upper bound for the frequency of the wave to overspin the black hole~\cite{Duztas:2019ick}.  However, if the initial black hole configuration is far from extremal and the wave mode satisfies $\omega < m \Omega_{\text{H}} $  (where $m\Omega_H\leq \omega_c $), after the absorption of the infinitesimal energy and angular momentum of the fields, the composite body would still be a black hole. From the first law of black hole thermodynamics, the horizon area would decrease,
\begin{equation}
\frac{1}{4}dA=\frac{dM}{T}-\frac{\Omega_H}{T}dJ=\frac{M_0 \epsilon}{T} \left( 1- \frac{m \Omega_H}{\omega}\right)<0.\label{AreaDec}
\end{equation}
This suggests that the assumption for the energy of the classical wave packet might be inappropriate.

%====================================
\section{Test fields and the weak cosmic censorship conjecture}\label{sec:3}
%====================================

In this section, we test the weak cosmic censorship conjecture for an extremal Kerr-Newman black hole with a test scalar field and a test neutrino field. Through the condition for overspin an extremal black hole, we explore the relationship between the weak cosmic censorship conjecture and the second law of black hole thermodynamics. 

%===================================
\subsection{Test scalar field and the weak cosmic censorship conjecture}
%===================================

The equation of motion for a test scalar field in the Kerr-Newman spacetime is governed by the Klein-Gordon equation, which is
\begin{equation}
	\frac{1}{\sqrt{-g}}\partial_{\mu}
 \left(\sqrt{-g}g^{\mu\nu}\partial_{\nu}\Psi\right)=0.\label{KG}
\end{equation}
Near the event horizon, the solution for the scalar field is
\begin{equation}
\Psi=\exp(-i\omega t)\exp\left[-i(\omega-m\Omega_{\text{H}})r_*\right]S_{lm}(\theta)\mathrm{e}^{im\phi},
\end{equation}
where $S_{lm}$ are spheroidal harmonics, and we have chosen the ingoing wave modes.
The energy-momentum tensor of the scalar field is
\begin{equation}\label{energy-momentun}
   T_{\mu\nu}=-\partial_{\left( \mu \right. } \Psi\partial_{\left. \nu \right)}\Psi^*
    +\frac{1}{2}g_{\mu\nu}\partial_\alpha\Psi\partial^\alpha\Psi^*.
\end{equation}
The energy and angular momentum fluxes through the event horizon are
\begin{align}
   \frac{dE}{dt}&=\int_\text{H} T^r_t\sqrt{-g} \, \mathrm{d}\theta \,\mathrm{d}\phi=\omega(\omega-m\Omega_{\text{H}})(r_\text{h}^2+a^2),\\
   \frac{dJ}{dt}&=-\int_\text{H} T^r_\phi\sqrt{-g} \, \mathrm{d}\theta \, \mathrm{d}\phi=m(\omega-m\Omega_{\text{H}})(r_\text{h}^2+a^2).
\end{align}

For a small time interval, the energy and angular momentum of the scalar field absorbed by the Kerr-Newman black hole are
\begin{align}
   dE&=\omega(\omega-m\Omega_{\text{H}})(r_\text{h}^2+a^2)dt,\label{energyflux}\\
   dJ&=m(\omega-m\Omega_{\text{H}})(r_\text{h}^2+a^2)dt.\label{angulamonentuflux}
\end{align}
According to the first law of black hole thermodynamics, after a Kerr-Newman black hole absorbs the scalar field, the change in the area of the black hole event horizon is
\begin{equation}
\begin{split}
dA&=\frac{4dM}{T}-\frac{4\Omega_H}{T}dJ\\
&=\frac{4}{T} \left( \omega-m \Omega_H\right)^2(\rh^2+a^2)dt\geq 0.
\end{split}
\end{equation}
The result suggests that the area of the black hole event horizon never decreases during the scattering of a scalar field, which is consistent with Hawking's area-increasing theorem.

Next, we examine the validity of the weak cosmic censorship conjecture during the scattering process. We consider an extremal Kerr-Newman black hole, and after absorbing the scalar field, the change in the minimum of the metric function of the resulting composite object is
\begin{equation}
\begin{split}
  d\Delta_{\text{min}}&= \frac{\partial\Delta_{\text{min}}}{\partial M}dM+\frac{\partial\Delta_{\text{min}}}{\partial J}dJ\\
   &=-\left(2M+\frac{2a^2}{M}\right)
    \left(\omega-\frac{ma}{M^2+a^2}\right)\\
    & \times\left(\omega-m\Omega_{\text{H}}\right)(\rh^2+a^2)dt.
\end{split}
\end{equation}
For an initial extremal Kerr-Newman black hole, after the scattering process, the minimum of the metric function becomes
\begin{equation}
    \begin{split}
       \Delta'_{\text{min}}=-\int  \left(2M+\frac{2a^2}{M}\right)\left(\omega-m\Omega_{\text{H}}\right)^2(\rh^2+a^2)\,\mathrm{d}t \leq 0.
    \end{split}
\end{equation}
The result shows that an extremal Kerr-Newman black hole cannot be destroyed by a test scalar field.

%===================================
\subsection{Test neutrino field and the weak cosmic censorship conjecture}
%===================================

The equation of motion for a neutrino field in the Kerr-Newman spacetime is governed by the Dirac equation. Although it can be derived using the Newman-Penrose spin-coefficient formalism~\cite{Page:1976jj,Lee:1977gk}, we choose to present it in the more conventional form.

A suitable choice of the $\gamma^\mu$ matrices, which satisfy the anticommutation relation $\{\gamma^\mu,\gamma^\nu\}=2g^{\mu\nu}$, is given by:
\begin{equation}
    \begin{split}
    \gamma^t&=\frac{r^2+a^2}{\sqrt{\Delta \Sigma}}\gamma^0+\frac{a \sin\theta}{\sqrt{\Sigma}
    }\gamma^2,\qquad
    \gamma^r=\sqrt{\frac{\Delta}{\Sigma}}\gamma^3,\\
    \gamma^\phi&=\frac{a}{\sqrt{\Delta\Sigma}}\gamma^0+\frac{1}{\sqrt{\Sigma}\sin\theta}\gamma^2,    \qquad \gamma^\theta=\frac{1}{\sqrt{\Sigma}}\gamma^1,
    \end{split}
\end{equation}
where $\gamma^a$'s refer to the standard Dirac matrices in flat spacetime, expressed in the Bjorken-Drell representation~\cite{Bjorken:1979dk}
%========================
\begin{equation}
\begin{split}
\gamma^0&=\begin{pmatrix}
 \text{I} & 0 \\
 0 & -\text{I} 
\end{pmatrix},\qquad \quad
\gamma^1=\begin{pmatrix}
 0 & \sigma^1 \\
 -\sigma^1 & 0 
\end{pmatrix},\\
\gamma^2&=\begin{pmatrix}
 0 & \sigma^2 \\
 -\sigma^2 & 0 
\end{pmatrix},\quad \quad
\gamma^3=\begin{pmatrix}
 0 & \sigma^3 \\
 -\sigma^3 & 0 
\end{pmatrix},
    \end{split}
\end{equation}
and the $\sigma$'s are the Pauli matrices
%================
\begin{equation}
    \begin{split}
        \text{I}&=
\begin{pmatrix}
 1 & 0 \\
 0 & 1 
 \end{pmatrix},\qquad \quad
\text{$\sigma^1 $}=
\begin{pmatrix}
 0 & 1 \\
 1 & 0 
\end{pmatrix},\\
\text{$\sigma^2 $}&=
\begin{pmatrix}
 0 & -i \\
 i & 0 
\end{pmatrix},
\qquad
\text{$\sigma^3 $}=
\begin{pmatrix}
 1 & 0 \\
 0 & -1
\end{pmatrix}. 
\end{split}
\end{equation}
%================
The resulting $\gamma^5$ is defined as 
\begin{equation}
\gamma^5=i\gamma^0\gamma^1\gamma^2\gamma^3.
\end{equation}
The equation of motion for a neutrino field is now given by:
\begin{equation}
    i\gamma^\mu\left(\partial_\mu-\Gamma_\mu\right)\Psi=0,
\end{equation}
where the $\Gamma_\mu$ are a set of spin-affine connection:
\begin{equation}
    \begin{split}
        \Gamma_\mu=\frac{i}{2}\Omega^{a~b}_{~\mu}\Sigma_{ab},\qquad \Sigma_{ab}=\frac{i}{4}\left[\gamma_a,\gamma_b\right].
    \end{split}
\end{equation}
To simplify our calculations and facilitate comparison with the neutrino field in the Kerr spacetime~\cite{Unruh:1973bda}, we consider solutions that satisfy $(1-\gamma^5)\Psi=0$. Then we have
\begin{equation}
\begin{split}
    \Psi=
\begin{pmatrix}
 \eta  \\
 \eta  
\end{pmatrix},
\end{split}
\end{equation}
where $\eta$ represents the two-component spinor, which can be further decomposed as follows:
\begin{equation}
\begin{split}
\eta=\frac{\mathrm{e}^{-i\omega t}\mathrm{e}^{-im\phi}}{\left[\Delta \sin^2\theta (r+ia\cos\theta)^2\right]^{1/4}}
\begin{pmatrix}
 R_1(r)S_1(\theta) \\
 R_2(r)S_2(\theta)  \\
\end{pmatrix}.
\end{split}
\end{equation}
Detailed calculations show that the equation of motion for the neutrino field can be reduced to the following four coupled first-order equations:
\begin{align}
     \begin{split}
        &\left(\frac{d}{dr}-i\omega \frac{r^2+a^2}{\Delta}-i\frac{ma}{\Delta}\right)R_1(r)=\frac{\lambda}{\Delta^{1/2}}R_2(r),\\
        &\left(\frac{d}{dr}+i\omega \frac{r^2+a^2}{\Delta}+i\frac{ma}{\Delta}\right)R_2(r)=\frac{\lambda}{\Delta^{1/2}}R_1(r);
    \end{split}\label{radiualEq} \\
     \begin{split}
        &\left(\frac{d}{d\theta}+\omega a \sin\theta+\frac{m}{\sin\theta}\right)S_1(\theta)=\lambda S_2(r),\\
         &\left(\frac{d}{d\theta}-\omega a \sin\theta-\frac{m}{\sin\theta}\right)S_2(\theta)=-\lambda S_1(r);
    \end{split}\label{angularEq}
\end{align}
where $\lambda$ is the separation constant determined by Eq.~\eqref{angularEq}. 
Interestingly, the equations of motion for the neutrino field in the Kerr-Newman spacetime take the same form as those in the Kerr spacetime~\cite{Unruh:1973bda}, with the only modification being the replacement of the metric function $\Delta$ with that of the Kerr-Newman spacetime. 

To solve the equation of motion for the neutrino field, we introduce the tortoise coordinate $r_*$, defined by:
\begin{equation}
    \frac{dr_*}{dr}=\frac{r^2+a^2}{\Delta}.
\end{equation}
Since we are more concerned with the radial equations, and the angular part of the field can be normalized, the radial equations~\eqref{radiualEq} can be simplified to
\begin{equation}
    \begin{split}
        \frac{dR_1}{dr_*}-i\left(\omega +\frac{ma}{r^2+a^2}\right)R_1&=\frac{\lambda\sqrt{\Delta}}{r^2+a^2}R_2,\\
        \frac{dR_2}{dr_*}+i\left(\omega +\frac{ma}{r^2+a^2}\right)R_2&=\frac{\lambda\sqrt{\Delta}}{r^2+a^2}R_1.
    \end{split}
\end{equation}
By considering the ingoing wave mode near the event horizon of the Kerr-Newman black hole, we have
\begin{equation}
    \begin{split}
        R_1(r) &\approx \mathcal{O}(\sqrt{\Delta}),\\
        R_2(r) &\approx \exp[-i(\omega+m\Omega_{\text{H}})r_*]+\mathcal{O}(\Delta).
    \end{split}
\end{equation}
Define the conserved neutrino number current as
\begin{equation}
    J^\mu (x)=\overline{\Psi}\gamma^\mu \Psi,
\end{equation}
where $\overline{\Psi}$ being the Dirac adjoint, $\Psi^{\dagger}\gamma^0$. The number current crossing the black hole event horizon is given by
\begin{equation}
    \begin{split}
        \frac{\partial N}{\partial t}&=-\int_{\text{H}} \,\mathrm{d}\theta\,\mathrm{ d}\phi\,\sqrt{-g}J^r \\
        &=-2\int_{\text{H}} \,\mathrm{d}\theta \,\mathrm{d}\phi \,\left[\abs{R_1(r)}^2\abs{S_1(\theta)}^2 -\abs{R_2(r)}^2\abs{S_2(\theta)}^2\right]\\
        &=-4\pi^2 \left[\abs{R_1(\rh)}^2 -\abs{R_2(\rh)}^2\right]\\
        &\approx 4\pi^2\abs{R_2(\rh)}^2>0.
    \end{split}
\end{equation}
Similar to the case for the scalar field, we have used the normalization condition for the angular functions
\begin{equation}
    \int \abs{S_1(\theta)}^2\,\mathrm{d}\theta \mathrm{d}\phi= \int \abs{S_2(\theta)}^2\,\mathrm{d}\theta\,\mathrm{d}\phi=1.
\end{equation}
The energy-momentum tensor for the neutrino field given by Brill and Wheeler is~\cite{Brill:1957fx}
\begin{equation}
    \begin{split}
        T_{\mu\nu}=\frac{i}{8\pi}\overline{\Psi}\left[\gamma_\mu\left(\partial_\nu-\Gamma_\nu\right)+\gamma_\nu\left(\partial_\mu-\Gamma_\mu\right)\right]\Psi+\text{c.c}.
    \end{split}
\end{equation}
From the energy-momentum tensor of the fermion, we can derive the energy and angular momentum fluxes of the test neutrino field, which are~\cite{Unruh:1973bda}
\begin{align}
   \frac{dE}{dt}&=\int_{\text{H}} T^r_t\sqrt{-g} \, \mathrm{d}\theta \,\mathrm{d}\phi=\omega\frac{\partial N}{\partial t},\\
   \frac{dJ}{dt}&=-\int_{\text{H}} T^r_\phi\sqrt{-g} \, \mathrm{d}\theta \,\mathrm{d}\phi=m\frac{\partial N}{\partial t}.
\end{align}
The results indicate that for fermion scattering, the energy and angular momentum absorbed by the black hole are always positive, and there is no supperradiance.

To use the black hole thermodynamical laws to check the change of the area of the black hole event horizon, we consider a Kerr-Newman black hole far from extremal. During the scattering process, the change in the area of the black hole event horizon is
\begin{equation}
\begin{split}
        dA&=\frac{4}{T}(dM-\Omega_{\text{H}}dJ)\\
    &=\frac{16\pi^2}{T}(\omega-m\Omega_{\text{H}})\abs{R_2(\rh)}^2dt.
\end{split}
\end{equation}
Evidently, for a neutrino field with frequency $\omega<m\Omega_{\text{H}}$, the black hole event horizon area decreases after absorbing the field. This decrease in the area of the event horizon during the scattering process violates Hawking's area-increasing theorem~\cite{Hawking:1971tu}.

By analyzing the changes in the black hole parameters during the scattering process, we can check the validity of the weak cosmic censorship conjecture. Considering an extremal black hole, we find that after the neutrino field being absorbed, the change in the minimum of the metric function is
\begin{equation}
    \begin{split}
        d\Delta_{\text{min}}&=\frac{\partial\Delta_{\text{min}}}{\partial M}dM+\frac{\partial\Delta_{\text{min}}}{\partial J}dJ\\
        &=-8\pi^2\left(M+\frac{a^2}{M}\right)\left(\omega-m\Omega_{\text{H}}\right)\abs{R(\rh)}^2dt.
    \end{split}
\end{equation}
After the neutrino field is absorbed by the black hole, the minimum value of the metric function for the extremal Kerr-Newman black hole is
\begin{equation}
    \begin{split}
        \Delta'_{\text{min}}=-8\pi^2\int \left(M+\frac{a^2}{M}\right)\left(\omega-m\Omega_{\text{H}}\right)\abs{R(\rh)}^2\,\mathrm{d}t.
    \end{split}
\end{equation}
For a neutrino field with frequency $\omega<m\Omega_{\text{H}}$, the minimum of the metric function is positive. This suggests that the event horizon of the extremal Kerr-Newman black hole disappears after the neutrino field is absorbed.

It should be noted that the frequency required to destroy the event horizon of the extremal Kerr-Newman black hole matches the frequency of the neutrino field that leads to the violation of Hawking’s area-increasing theorem. This suggests a profound connection between the second law of black hole thermodynamics and the weak cosmic censorship conjecture.

%====================================
\section{The second law and the cosmic censor}\label{Sec:general}
%====================================

As demonstrated in Sec.~\ref{sec:2b}, the findings of previous research imply that a Kerr-Newman black hole may be destroyed by a test scalar field and a test neutrino field. Consequently, this suggests that the weak cosmic censorship does not hold for the Kerr-Newman black hole. Our analysis indicates that the assumption of the energy and angular momentum absorbed by the Kerr-Newman black hole in the previous studies violates the second law of black hole thermodynamics. As our work suggested in the previous section, the energy and angular momentum of the test scalar field absorbed by an extremal black hole satisfies the second law of black hole thermodynamics and cannot destroy an extremal Kerr-Newman black hole. However, in the case of the neutrino field scattering, not only the energy and angular momentum of the neutrino field absorbed by the extremal black hole violate the second law of black hole thermodynamics, but also the event horizon of the extremal black hole can be destroyed. Furthermore, the condition to violate the second law of black hole thermodynamics is the same as the condition to destroy an extremal Kerr-Newman black hole. It intuitively suggests that the second law of black hole thermodynamics guarantees the weak cosmic censorship conjecture and the violation of the second law inevitably leads to the violation of the weak cosmic censorship conjecture.

% It seems that the argument is trivial since the thermodynamics for a Kerr-Newman black hole is well developed. 
Consider a more subtle case for the Gedanken experiment to destroy the event horizon for black holes that the thermodynamical laws are not well established. Black holes with NUT parameter in general relativity are the cases. There are several kinds of different viewpoints regarding black holes with NUT parameters~\cite{Hennigar:2019ive,Bordo:2019tyh,Chen:2019uhp,Wu:2019pzr,Liu:2022wku} and consensus about the thermodynamics of black holes with NUT parameters is still beyond reach. Consider destroying the event horizon of a Kerr Taub-NUT black hole with a test scalar field. If we use inappropriate thermodynamics, a test field can destroy the extremal Kerr Taub-NUT black hole, as the study~\cite{Duztas:2017lxk} indicated. However, it also violates the second law of black hole thermodynamics. When appropriate thermodynamics is considered, our previous work~\cite{Yang:2020iat} suggested that the extremal Kerr Taub-NUT black hole cannot be destroyed by a test scalar field, and the second law of black hole thermodynamics holds.

Additionally, for other black holes such as the $3$D Ba\~nados–Teitelboim–Zanelli (BTZ) black hole in general relativity and the Kerr-Sen black hole in modified theory of gravity, it was claimed that these black holes can also be destroyed by test fields~\cite{Duztas:2016xfg,Duztas:2018adf}. However, later works suggested that neither the BTZ black hole nor the Kerr-Sen black hole can be destroyed by a complex test scalar field~\cite{Chen:2018yah,Gwak:2019rcz}. 
The essential difference between the results of later studies and the results of the early work for the BTZ and Kerr-Sen black holes is the different choice of the energy and angular momentum of the incoming wave packet, and hence the change of the black hole parameters. The choice of the energy and angular momentum of the scalar field in later studies is based on the energy-momentum tensor of the scalar field, and the changes of the black hole parameters are derived from the fluxes of the test fields across the event horizon. They are consistent with Hawking's area-increasing theorem. However, the assumption of early works for the energy and angular momentum fluxes for the BTZ and Kerr-Sen black holes also violates Hawking's area-increasing theorem. Consequently, this is consistent with the result of violating the weak cosmic censorship conjecture.

As we noticed in the previous section, the energy and angular momentum of the test scalar field absorbed by an extremal black hole satisfy the second law of black hole thermodynamics and cannot destroy an extremal Kerr-Newman black hole. However, for neutrino field scattering, the energy and angular momentum absorbed by the extremal black hole not only violate the second law of black hole thermodynamics but also can destroy the event horizon of the extremal black hole. Furthermore, the condition to violate the second law of black hole thermodynamics is the same as the condition to destroy an extremal Kerr-Newman black hole.

We considered testing the weak cosmic censorship conjecture from classical viewpoints without quantizing the test fields. We treated the test scalar field and neutrino field as classical fields. For a scalar field, the energy-momentum tensor satisfies the null energy condition. However, the neutrino field does not satisfy the null energy condition~\cite{Toth:2015cda}. Furthermore, there is no classical Fermi field~\cite{Unruh:1973bda}, Hawking's area-increasing theorem and the weak cosmic censorship conjecture might only be true classically. The scattering of the Fermi field might inevitably lead to the violation of the second law of black hole thermodynamics and hence the weak cosmic censorship conjecture.

Consistent with, yet distinct from, Toth's perspective on the null energy condition in investigation for a test charged scalar field and a Dirac field for checking the weak cosmic censorship conjecture for a dyonic Kerr-Newman black hole~\cite{Toth:2012vvy,Toth:2015cda}, we present our arguments from the relationship between the second law of black hole thermodynamics and the weak cosmic censorship conjecture. Our viewpoint is consistent with the recent study by Wu et al.~\cite{Wu:2024ucf,Lu:2025ntu}. In their work, they defined a parameter $w=\left(\frac{\partial S}{\partial T}\right)_{Q_i;T=0}$, and gave a general mathematical proof that the validity of the weak cosmic censorship conjecture for the extremal black hole depends on the sign of the parameter $w$. They found that $w>0$ preserves the weak cosmic censorship conjecture and $w<0$ would indicate a potential violation~\cite{Wu:2024ucf}. Our work seems to give a physical viewpoint for the relation between the weak cosmic censorship conjecture and the second law of black hole thermodynamics.

%====================================
\section{Conclusion and discussion}\label{sec:C&D}
%====================================

Both the weak cosmic censorship conjecture and black hole thermodynamics are the foundations for black hole physics. The violation of the weak cosmic censorship conjecture would lead to the lost predictability of gravitational theory. In this paper, we reexamined the validity of the weak cosmic censorship conjecture for a Kerr-Newman black hole, and explored its relationship with the second law of black hole thermodynamics. In the previous work, the author claimed that a Kerr-Newman black hole can be generally overspun and argued that even if the backreactions are taken into account, it can still be overspun. The author concluded that the weak cosmic censorship conjecture can be violated for a Kerr-Newman black hole. In our paper, we pointed out that the assumption for the energy and angular momentum of the test field in the previous work violates the second law of black hole thermodynamics and demonstrated that an extremal Kerr-Newman black hole cannot be overspun by a test scalar field but can be destroyed by a neutrino field. We found that the condition to overspin an extremal Kerr-Newman black hole is the same as the condition to violate the second law, and argued that the violation of the second law of black hole thermodynamics might inevitably lead to a violation of the weak cosmic censorship conjecture.

In the gedanken experiment to destroy a black hole, the focus lies solely on the nature of the final state—whether it remains a black hole or becomes a naked singularity—while the intermediate process is disregarded. In particular, we neglect the backreaction of the matter field on the spacetime before it is absorbed by the black hole. The assumption that the final composite object can be described by the Kerr–Newman metric is based on the black hole no-hair theorem, which states that any stationary, asymptotically flat, and analytic solution to the Einstein–Maxwell equations is uniquely characterized by its mass, angular momentum, and electric charge~\cite{Misner:1973prb}.

In this work, we focus on single-mode field configurations with monotonic profiles, which allow for a clearer analysis of the scattering process. For incident matter fields with nontrivial angular profiles, a decomposition into individual modes $S_{lm}(\theta)$ can be performed. One can show that the conclusions remain unchanged, analogous to the analysis conducted for the weak cosmic censorship conjecture in the context of Kerr–Newman Taub–NUT black holes using non-monotonic fields~\cite{Yang:2023hll}.
When the matter field is time-dependent, the scattering process can be further divided into a sequence of infinitesimally short time intervals. Each interval can be treated independently by updating the initial values of black hole parameters accordingly. This method reproduces the same conclusions~\cite{Yang:2023hll}.

Through the scattering of scalar and neutrino fields, we show that when an infalling matter field with low energy density and angular momentum reduces the horizon area of a non-extremal black hole, the same perturbation would result in the destruction of the horizon in the extremal case. It is particularly intriguing to consider the possibility of establishing this relationship in full generality, without assuming any details of the infalling matter.

%=====================================
\acknowledgments

We thank Wen-Di Guo, H. Khodabakhshi and Shan-Ping Wu for useful discussions. This work was supported by the National Natural Science Foundation of China (Grants No. 12305065, No. 12475056,  No. 12475055 and No. 12247101), the China Postdoctoral Science Foundation (Grant No. 2023M731468), the Gansu Province's Top Leading Talent Support Plan, the Fundamental Research Funds for the Central Universities (Grant No. lzujbky-2024-jdzx06), the Natural Science Foundation of Gansu Province (No. 22JR5RA389), and the `111 Center' under Grant No. B20063.
%=====================================
%===========================================
% \textbf{Foundation:}\\
% YSJ:\\
% NSFC: No. 12305065 (2024.01--2026.12)  \qquad  No. 12247178 (2023.01--2023.12)\\
% the China Postdoctoral Science Foundation: Grant No. 2023M731468    \\
% Ali:  NSFC: No. 12347177 (2024.01--2024.12)\\
% WSW: \\
% NSFC:  Grants No. 12475055.\\
% LYX:\\
% National Key Research and Development Program of China: Grant No. 2021YFC2203003 \\
% LZ Center:\\
% NSFC: Grant No. 12247101
%===========================================
%

%=====================================
%\bibliographystyle{utphysmod}
%\bibliography{ref}

\begin{thebibliography}{72}%
\makeatletter
\providecommand \@ifxundefined [1]{%
 \@ifx{#1\undefined}
}%
\providecommand \@ifnum [1]{%
 \ifnum #1\expandafter \@firstoftwo
 \else \expandafter \@secondoftwo
 \fi
}%
\providecommand \@ifx [1]{%
 \ifx #1\expandafter \@firstoftwo
 \else \expandafter \@secondoftwo
 \fi
}%
\providecommand \natexlab [1]{#1}%
\providecommand \enquote  [1]{``#1''}%
\providecommand \bibnamefont  [1]{#1}%
\providecommand \bibfnamefont [1]{#1}%
\providecommand \citenamefont [1]{#1}%
\providecommand \href@noop [0]{\@secondoftwo}%
\providecommand \href [0]{\begingroup \@sanitize@url \@href}%
\providecommand \@href[1]{\@@startlink{#1}\@@href}%
\providecommand \@@href[1]{\endgroup#1\@@endlink}%
\providecommand \@sanitize@url [0]{\catcode `\\12\catcode `\$12\catcode
  `\&12\catcode `\#12\catcode `\^12\catcode `\_12\catcode `\%12\relax}%
\providecommand \@@startlink[1]{}%
\providecommand \@@endlink[0]{}%
\providecommand \url  [0]{\begingroup\@sanitize@url \@url }%
\providecommand \@url [1]{\endgroup\@href {#1}{\urlprefix }}%
\providecommand \urlprefix  [0]{URL }%
\providecommand \Eprint [0]{\href }%
\providecommand \doibase [0]{https://doi.org/}%
\providecommand \selectlanguage [0]{\@gobble}%
\providecommand \bibinfo  [0]{\@secondoftwo}%
\providecommand \bibfield  [0]{\@secondoftwo}%
\providecommand \translation [1]{[#1]}%
\providecommand \BibitemOpen [0]{}%
\providecommand \bibitemStop [0]{}%
\providecommand \bibitemNoStop [0]{.\EOS\space}%
\providecommand \EOS [0]{\spacefactor3000\relax}%
\providecommand \BibitemShut  [1]{\csname bibitem#1\endcsname}%
\let\auto@bib@innerbib\@empty
%</preamble>
\bibitem [{\citenamefont {Ruffini}\ and\ \citenamefont
  {Wheeler}(1971)}]{Ruffini:1971bza}%
  \BibitemOpen
  \bibfield  {author} {\bibinfo {author} {\bibfnamefont {R.}~\bibnamefont
  {Ruffini}}\ and\ \bibinfo {author} {\bibfnamefont {J.~A.}\ \bibnamefont
  {Wheeler}},\ }\bibfield  {title} {\bibinfo {title} {{Introducing the black
  hole}},\ }\href {https://doi.org/10.1063/1.3022513} {\bibfield  {journal}
  {\bibinfo  {journal} {Phys. Today}\ }\textbf {\bibinfo {volume} {24}},\
  \bibinfo {pages} {30} (\bibinfo {year} {1971})}\BibitemShut {NoStop}%
\bibitem [{\citenamefont {Penrose}(1969)}]{Penr69}%
  \BibitemOpen
  \bibfield  {author} {\bibinfo {author} {\bibfnamefont {R.}~\bibnamefont
  {Penrose}},\ }\bibfield  {title} {\bibinfo {title} {{Gravitational collapse:
  The role of general relativity}},\ }\href
  {https://doi.org/10.1023/A:1016578408204} {\bibfield  {journal} {\bibinfo
  {journal} {Riv. Nuovo Cim.}\ }\textbf {\bibinfo {volume} {1}},\ \bibinfo
  {pages} {252} (\bibinfo {year} {1969})}\BibitemShut {NoStop}%
\bibitem [{\citenamefont {Bardeen}\ \emph {et~al.}(1973)\citenamefont
  {Bardeen}, \citenamefont {Carter},\ and\ \citenamefont
  {Hawking}}]{Bardeen:1973gs}%
  \BibitemOpen
  \bibfield  {author} {\bibinfo {author} {\bibfnamefont {J.~M.}\ \bibnamefont
  {Bardeen}}, \bibinfo {author} {\bibfnamefont {B.}~\bibnamefont {Carter}},\
  and\ \bibinfo {author} {\bibfnamefont {S.~W.}\ \bibnamefont {Hawking}},\
  }\bibfield  {title} {\bibinfo {title} {{The Four laws of black hole
  mechanics}},\ }\href {https://doi.org/10.1007/BF01645742} {\bibfield
  {journal} {\bibinfo  {journal} {Commun. Math. Phys.}\ }\textbf {\bibinfo
  {volume} {31}},\ \bibinfo {pages} {161} (\bibinfo {year} {1973})}\BibitemShut
  {NoStop}%
\bibitem [{\citenamefont {Bekenstein}(1973{\natexlab{a}})}]{Bekenstein:1973ur}%
  \BibitemOpen
  \bibfield  {author} {\bibinfo {author} {\bibfnamefont {J.~D.}\ \bibnamefont
  {Bekenstein}},\ }\bibfield  {title} {\bibinfo {title} {{Black holes and
  entropy}},\ }\href {https://doi.org/10.1103/PhysRevD.7.2333} {\bibfield
  {journal} {\bibinfo  {journal} {Phys. Rev. D}\ }\textbf {\bibinfo {volume}
  {7}},\ \bibinfo {pages} {2333} (\bibinfo {year}
  {1973}{\natexlab{a}})}\BibitemShut {NoStop}%
\bibitem [{\citenamefont {Christodoulou}(1984)}]{Chri84}%
  \BibitemOpen
  \bibfield  {author} {\bibinfo {author} {\bibfnamefont {D.}~\bibnamefont
  {Christodoulou}},\ }\bibfield  {title} {\bibinfo {title} {{Violation of
  cosmic censorship in the gravitational collapse of a dust cloud}},\ }\href
  {https://doi.org/10.1007/BF01223743} {\bibfield  {journal} {\bibinfo
  {journal} {Commun. Math. Phys.}\ }\textbf {\bibinfo {volume} {93}},\ \bibinfo
  {pages} {171} (\bibinfo {year} {1984})}\BibitemShut {NoStop}%
\bibitem [{\citenamefont {Ori}\ and\ \citenamefont {Piran}(1987)}]{OrPi87}%
  \BibitemOpen
  \bibfield  {author} {\bibinfo {author} {\bibfnamefont {A.}~\bibnamefont
  {Ori}}\ and\ \bibinfo {author} {\bibfnamefont {T.}~\bibnamefont {Piran}},\
  }\bibfield  {title} {\bibinfo {title} {{Naked Singularities in Selfsimilar
  Spherical Gravitational Collapse}},\ }\href
  {https://doi.org/10.1103/PhysRevLett.59.2137} {\bibfield  {journal} {\bibinfo
   {journal} {Phys. Rev. Lett.}\ }\textbf {\bibinfo {volume} {59}},\ \bibinfo
  {pages} {2137} (\bibinfo {year} {1987})}\BibitemShut {NoStop}%
\bibitem [{\citenamefont {Lemos}(1992)}]{Lemo92}%
  \BibitemOpen
  \bibfield  {author} {\bibinfo {author} {\bibfnamefont {J.~P.~S.}\
  \bibnamefont {Lemos}},\ }\bibfield  {title} {\bibinfo {title} {{Naked
  singularities: Gravitationally collapsing configurations of dust or radiation
  in spherical symmetry. A Unified treatment}},\ }\href
  {https://doi.org/10.1103/PhysRevLett.68.1447} {\bibfield  {journal} {\bibinfo
   {journal} {Phys. Rev. Lett.}\ }\textbf {\bibinfo {volume} {68}},\ \bibinfo
  {pages} {1447} (\bibinfo {year} {1992})}\BibitemShut {NoStop}%
\bibitem [{\citenamefont {Corelli}\ \emph {et~al.}(2021)\citenamefont
  {Corelli}, \citenamefont {Ikeda},\ and\ \citenamefont
  {Pani}}]{Corelli:2021ikv}%
  \BibitemOpen
  \bibfield  {author} {\bibinfo {author} {\bibfnamefont {F.}~\bibnamefont
  {Corelli}}, \bibinfo {author} {\bibfnamefont {T.}~\bibnamefont {Ikeda}},\
  and\ \bibinfo {author} {\bibfnamefont {P.}~\bibnamefont {Pani}},\ }\bibfield
  {title} {\bibinfo {title} {{Challenging cosmic censorship in
  Einstein-Maxwell-scalar theory with numerically simulated gedanken
  experiments}},\ }\href {https://doi.org/10.1103/PhysRevD.104.084069}
  {\bibfield  {journal} {\bibinfo  {journal} {Phys. Rev. D}\ }\textbf {\bibinfo
  {volume} {104}},\ \bibinfo {pages} {084069} (\bibinfo {year} {2021})},\
  \Eprint {https://arxiv.org/abs/2108.08328} {arXiv:2108.08328 [gr-qc]}
  \BibitemShut {NoStop}%
\bibitem [{\citenamefont {Eperon}\ \emph {et~al.}(2020)\citenamefont {Eperon},
  \citenamefont {Ganchev},\ and\ \citenamefont {Santos}}]{Eperon:2019viw}%
  \BibitemOpen
  \bibfield  {author} {\bibinfo {author} {\bibfnamefont {F.~C.}\ \bibnamefont
  {Eperon}}, \bibinfo {author} {\bibfnamefont {B.}~\bibnamefont {Ganchev}},\
  and\ \bibinfo {author} {\bibfnamefont {J.~E.}\ \bibnamefont {Santos}},\
  }\bibfield  {title} {\bibinfo {title} {{Plausible scenario for a generic
  violation of the weak cosmic censorship conjecture in asymptotically flat
  four dimensions}},\ }\href {https://doi.org/10.1103/PhysRevD.101.041502}
  {\bibfield  {journal} {\bibinfo  {journal} {Phys. Rev. D}\ }\textbf {\bibinfo
  {volume} {101}},\ \bibinfo {pages} {041502} (\bibinfo {year} {2020})},\
  \Eprint {https://arxiv.org/abs/1906.11257} {arXiv:1906.11257 [gr-qc]}
  \BibitemShut {NoStop}%
\bibitem [{\citenamefont {Hu}\ \emph {et~al.}(2020)\citenamefont {Hu},
  \citenamefont {Song}, \citenamefont {Sun}, \citenamefont {Li},\ and\
  \citenamefont {Wang}}]{HSSLW20}%
  \BibitemOpen
  \bibfield  {author} {\bibinfo {author} {\bibfnamefont {T.-T.}\ \bibnamefont
  {Hu}}, \bibinfo {author} {\bibfnamefont {Y.}~\bibnamefont {Song}}, \bibinfo
  {author} {\bibfnamefont {S.}~\bibnamefont {Sun}}, \bibinfo {author}
  {\bibfnamefont {H.-B.}\ \bibnamefont {Li}},\ and\ \bibinfo {author}
  {\bibfnamefont {Y.-Q.}\ \bibnamefont {Wang}},\ }\bibfield  {title} {\bibinfo
  {title} {{Weak cosmic censorship in Born\textendash{}Infeld electrodynamics
  and bound on charge-to-mass ratio}},\ }\href
  {https://doi.org/10.1140/epjc/s10052-020-7703-6} {\bibfield  {journal}
  {\bibinfo  {journal} {Eur. Phys. J. C}\ }\textbf {\bibinfo {volume} {80}},\
  \bibinfo {pages} {147} (\bibinfo {year} {2020})},\ \Eprint
  {https://arxiv.org/abs/1906.00235} {arXiv:1906.00235 [hep-th]} \BibitemShut
  {NoStop}%
\bibitem [{\citenamefont {Song}\ \emph {et~al.}(2024)\citenamefont {Song},
  \citenamefont {Cui},\ and\ \citenamefont {Wang}}]{Song:2022dpw}%
  \BibitemOpen
  \bibfield  {author} {\bibinfo {author} {\bibfnamefont {Y.}~\bibnamefont
  {Song}}, \bibinfo {author} {\bibfnamefont {S.-Y.}\ \bibnamefont {Cui}},\ and\
  \bibinfo {author} {\bibfnamefont {Y.-Q.}\ \bibnamefont {Wang}},\ }\bibfield
  {title} {\bibinfo {title} {{Weak cosmic censorship with SU(2) gauge field and
  bound on charge-to-mass ratio}},\ }\href
  {https://doi.org/10.1007/JHEP01(2024)100} {\bibfield  {journal} {\bibinfo
  {journal} {JHEP}\ }\textbf {\bibinfo {volume} {01}},\ \bibinfo {pages}
  {100}},\ \Eprint {https://arxiv.org/abs/2207.11703} {arXiv:2207.11703
  [hep-th]} \BibitemShut {NoStop}%
\bibitem [{\citenamefont {Cui}\ \emph {et~al.}(2024)\citenamefont {Cui},
  \citenamefont {Fang},\ and\ \citenamefont {Wang}}]{Cui:2024xqg}%
  \BibitemOpen
  \bibfield  {author} {\bibinfo {author} {\bibfnamefont {S.-Y.}\ \bibnamefont
  {Cui}}, \bibinfo {author} {\bibfnamefont {T.-F.}\ \bibnamefont {Fang}},\ and\
  \bibinfo {author} {\bibfnamefont {Y.-Q.}\ \bibnamefont {Wang}},\ }\bibfield
  {title} {\bibinfo {title} {{Weak cosmic censorship with excited scalar fields
  and bound on charge-to-mass ratio}},\ }\href
  {https://doi.org/10.1007/JHEP11(2024)085} {\bibfield  {journal} {\bibinfo
  {journal} {JHEP}\ }\textbf {\bibinfo {volume} {11}},\ \bibinfo {pages}
  {085}},\ \Eprint {https://arxiv.org/abs/2401.07866} {arXiv:2401.07866
  [hep-th]} \BibitemShut {NoStop}%
\bibitem [{\citenamefont {Sperhake}\ \emph {et~al.}(2009)\citenamefont
  {Sperhake}, \citenamefont {Cardoso}, \citenamefont {Pretorius}, \citenamefont
  {Berti}, \citenamefont {Hinderer},\ and\ \citenamefont {Yunes}}]{SCPBHY09}%
  \BibitemOpen
  \bibfield  {author} {\bibinfo {author} {\bibfnamefont {U.}~\bibnamefont
  {Sperhake}}, \bibinfo {author} {\bibfnamefont {V.}~\bibnamefont {Cardoso}},
  \bibinfo {author} {\bibfnamefont {F.}~\bibnamefont {Pretorius}}, \bibinfo
  {author} {\bibfnamefont {E.}~\bibnamefont {Berti}}, \bibinfo {author}
  {\bibfnamefont {T.}~\bibnamefont {Hinderer}},\ and\ \bibinfo {author}
  {\bibfnamefont {N.}~\bibnamefont {Yunes}},\ }\bibfield  {title} {\bibinfo
  {title} {{Cross section, final spin and zoom-whirl behavior in high-energy
  black hole collisions}},\ }\href
  {https://doi.org/10.1103/PhysRevLett.103.131102} {\bibfield  {journal}
  {\bibinfo  {journal} {Phys. Rev. Lett.}\ }\textbf {\bibinfo {volume} {103}},\
  \bibinfo {pages} {131102} (\bibinfo {year} {2009})},\ \Eprint
  {https://arxiv.org/abs/0907.1252} {arXiv:0907.1252 [gr-qc]} \BibitemShut
  {NoStop}%
\bibitem [{\citenamefont {Andrade}\ \emph
  {et~al.}(2019{\natexlab{a}})\citenamefont {Andrade}, \citenamefont {Emparan},
  \citenamefont {Licht},\ and\ \citenamefont {Luna}}]{AELL19}%
  \BibitemOpen
  \bibfield  {author} {\bibinfo {author} {\bibfnamefont {T.}~\bibnamefont
  {Andrade}}, \bibinfo {author} {\bibfnamefont {R.}~\bibnamefont {Emparan}},
  \bibinfo {author} {\bibfnamefont {D.}~\bibnamefont {Licht}},\ and\ \bibinfo
  {author} {\bibfnamefont {R.}~\bibnamefont {Luna}},\ }\bibfield  {title}
  {\bibinfo {title} {{Black hole collisions, instabilities, and cosmic
  censorship violation at large $D$}},\ }\href
  {https://doi.org/10.1007/JHEP09(2019)099} {\bibfield  {journal} {\bibinfo
  {journal} {JHEP}\ }\textbf {\bibinfo {volume} {09}},\ \bibinfo {pages}
  {099}},\ \Eprint {https://arxiv.org/abs/1908.03424} {arXiv:1908.03424
  [hep-th]} \BibitemShut {NoStop}%
\bibitem [{\citenamefont {Andrade}\ \emph
  {et~al.}(2019{\natexlab{b}})\citenamefont {Andrade}, \citenamefont {Emparan},
  \citenamefont {Licht},\ and\ \citenamefont {Luna}}]{AELL19a}%
  \BibitemOpen
  \bibfield  {author} {\bibinfo {author} {\bibfnamefont {T.}~\bibnamefont
  {Andrade}}, \bibinfo {author} {\bibfnamefont {R.}~\bibnamefont {Emparan}},
  \bibinfo {author} {\bibfnamefont {D.}~\bibnamefont {Licht}},\ and\ \bibinfo
  {author} {\bibfnamefont {R.}~\bibnamefont {Luna}},\ }\bibfield  {title}
  {\bibinfo {title} {{Cosmic censorship violation in black hole collisions in
  higher dimensions}},\ }\href {https://doi.org/10.1007/JHEP04(2019)121}
  {\bibfield  {journal} {\bibinfo  {journal} {JHEP}\ }\textbf {\bibinfo
  {volume} {04}},\ \bibinfo {pages} {121}},\ \Eprint
  {https://arxiv.org/abs/1812.05017} {arXiv:1812.05017 [hep-th]} \BibitemShut
  {NoStop}%
\bibitem [{\citenamefont {Wald}(1974)}]{Wald74}%
  \BibitemOpen
  \bibfield  {author} {\bibinfo {author} {\bibfnamefont {R.}~\bibnamefont
  {Wald}},\ }\bibfield  {title} {\bibinfo {title} {Gedanken experiments to
  destroy a black hole},\ }\href
  {https://doi.org/https://doi.org/10.1016/0003-4916(74)90125-0} {\bibfield
  {journal} {\bibinfo  {journal} {Annals of Physics}\ }\textbf {\bibinfo
  {volume} {82}},\ \bibinfo {pages} {548} (\bibinfo {year} {1974})}\BibitemShut
  {NoStop}%
\bibitem [{\citenamefont {Hubeny}(1999)}]{Hube99}%
  \BibitemOpen
  \bibfield  {author} {\bibinfo {author} {\bibfnamefont {V.~E.}\ \bibnamefont
  {Hubeny}},\ }\bibfield  {title} {\bibinfo {title} {{Overcharging a black hole
  and cosmic censorship}},\ }\href {https://doi.org/10.1103/PhysRevD.59.064013}
  {\bibfield  {journal} {\bibinfo  {journal} {Phys. Rev. D}\ }\textbf {\bibinfo
  {volume} {59}},\ \bibinfo {pages} {064013} (\bibinfo {year} {1999})},\
  \Eprint {https://arxiv.org/abs/gr-qc/9808043} {arXiv:gr-qc/9808043}
  \BibitemShut {NoStop}%
\bibitem [{\citenamefont {Hod}(2002)}]{Hod02}%
  \BibitemOpen
  \bibfield  {author} {\bibinfo {author} {\bibfnamefont {S.}~\bibnamefont
  {Hod}},\ }\bibfield  {title} {\bibinfo {title} {{Cosmic censorship, area
  theorem, and selfenergy of particles}},\ }\href
  {https://doi.org/10.1103/PhysRevD.66.024016} {\bibfield  {journal} {\bibinfo
  {journal} {Phys. Rev. D}\ }\textbf {\bibinfo {volume} {66}},\ \bibinfo
  {pages} {024016} (\bibinfo {year} {2002})},\ \Eprint
  {https://arxiv.org/abs/gr-qc/0205005} {arXiv:gr-qc/0205005} \BibitemShut
  {NoStop}%
\bibitem [{\citenamefont {Jacobson}\ and\ \citenamefont
  {Sotiriou}(2009)}]{JaSo09}%
  \BibitemOpen
  \bibfield  {author} {\bibinfo {author} {\bibfnamefont {T.}~\bibnamefont
  {Jacobson}}\ and\ \bibinfo {author} {\bibfnamefont {T.~P.}\ \bibnamefont
  {Sotiriou}},\ }\bibfield  {title} {\bibinfo {title} {{Over-spinning a black
  hole with a test body}},\ }\href
  {https://doi.org/10.1103/PhysRevLett.103.141101} {\bibfield  {journal}
  {\bibinfo  {journal} {Phys. Rev. Lett.}\ }\textbf {\bibinfo {volume} {103}},\
  \bibinfo {pages} {141101} (\bibinfo {year} {2009})},\ \bibinfo {note}
  {[Erratum: Phys.Rev.Lett. 103, 209903 (2009)]},\ \Eprint
  {https://arxiv.org/abs/0907.4146} {arXiv:0907.4146 [gr-qc]} \BibitemShut
  {NoStop}%
\bibitem [{\citenamefont {Yang}\ \emph {et~al.}(2022)\citenamefont {Yang},
  \citenamefont {Zhang}, \citenamefont {Wei},\ and\ \citenamefont
  {Liu}}]{Yang:2022yvq}%
  \BibitemOpen
  \bibfield  {author} {\bibinfo {author} {\bibfnamefont {S.-J.}\ \bibnamefont
  {Yang}}, \bibinfo {author} {\bibfnamefont {Y.-P.}\ \bibnamefont {Zhang}},
  \bibinfo {author} {\bibfnamefont {S.-W.}\ \bibnamefont {Wei}},\ and\ \bibinfo
  {author} {\bibfnamefont {Y.-X.}\ \bibnamefont {Liu}},\ }\bibfield  {title}
  {\bibinfo {title} {{Destroying the event horizon of a nonsingular rotating
  quantum-corrected black hole}},\ }\href
  {https://doi.org/10.1007/JHEP04(2022)066} {\bibfield  {journal} {\bibinfo
  {journal} {JHEP}\ }\textbf {\bibinfo {volume} {04}},\ \bibinfo {pages}
  {066}},\ \Eprint {https://arxiv.org/abs/2201.03381} {arXiv:2201.03381
  [gr-qc]} \BibitemShut {NoStop}%
\bibitem [{\citenamefont {Ghosh}\ \emph {et~al.}(2019)\citenamefont {Ghosh},
  \citenamefont {Fairoos},\ and\ \citenamefont {Sarkar}}]{Ghosh:2019dzq}%
  \BibitemOpen
  \bibfield  {author} {\bibinfo {author} {\bibfnamefont {R.}~\bibnamefont
  {Ghosh}}, \bibinfo {author} {\bibfnamefont {C.}~\bibnamefont {Fairoos}},\
  and\ \bibinfo {author} {\bibfnamefont {S.}~\bibnamefont {Sarkar}},\
  }\bibfield  {title} {\bibinfo {title} {{Overcharging higher curvature black
  holes}},\ }\href {https://doi.org/10.1103/PhysRevD.100.124019} {\bibfield
  {journal} {\bibinfo  {journal} {Phys. Rev. D}\ }\textbf {\bibinfo {volume}
  {100}},\ \bibinfo {pages} {124019} (\bibinfo {year} {2019})},\ \Eprint
  {https://arxiv.org/abs/1906.08016} {arXiv:1906.08016 [gr-qc]} \BibitemShut
  {NoStop}%
\bibitem [{\citenamefont {Liu}\ and\ \citenamefont {Gao}(2020)}]{Liu:2020cji}%
  \BibitemOpen
  \bibfield  {author} {\bibinfo {author} {\bibfnamefont {C.}~\bibnamefont
  {Liu}}\ and\ \bibinfo {author} {\bibfnamefont {S.}~\bibnamefont {Gao}},\
  }\bibfield  {title} {\bibinfo {title} {{Overcharging magnetized black holes
  at linear order and the weak cosmic censorship conjecture}},\ }\href
  {https://doi.org/10.1103/PhysRevD.101.124067} {\bibfield  {journal} {\bibinfo
   {journal} {Phys. Rev. D}\ }\textbf {\bibinfo {volume} {101}},\ \bibinfo
  {pages} {124067} (\bibinfo {year} {2020})},\ \Eprint
  {https://arxiv.org/abs/2003.12999} {arXiv:2003.12999 [gr-qc]} \BibitemShut
  {NoStop}%
\bibitem [{\citenamefont {Miao}\ and\ \citenamefont
  {Yang}(2025)}]{Miao:2024wwd}%
  \BibitemOpen
  \bibfield  {author} {\bibinfo {author} {\bibfnamefont {W.-J.}\ \bibnamefont
  {Miao}}\ and\ \bibinfo {author} {\bibfnamefont {S.-J.}\ \bibnamefont
  {Yang}},\ }\bibfield  {title} {\bibinfo {title} {{Overcharging a nonsingular
  black hole in general relativity: the nonlinear electrodynamic field
  effects}},\ }\href {https://doi.org/10.1088/1475-7516/2025/05/022} {\bibfield
   {journal} {\bibinfo  {journal} {JCAP}\ }\textbf {\bibinfo {volume} {05}},\
  \bibinfo {pages} {022}},\ \Eprint {https://arxiv.org/abs/2409.07305}
  {arXiv:2409.07305 [gr-qc]} \BibitemShut {NoStop}%
\bibitem [{\citenamefont {Gao}\ and\ \citenamefont {Zhang}(2013)}]{Gao:2012ca}%
  \BibitemOpen
  \bibfield  {author} {\bibinfo {author} {\bibfnamefont {S.}~\bibnamefont
  {Gao}}\ and\ \bibinfo {author} {\bibfnamefont {Y.}~\bibnamefont {Zhang}},\
  }\bibfield  {title} {\bibinfo {title} {{Destroying extremal Kerr-Newman black
  holes with test particles}},\ }\href
  {https://doi.org/10.1103/PhysRevD.87.044028} {\bibfield  {journal} {\bibinfo
  {journal} {Phys. Rev. D}\ }\textbf {\bibinfo {volume} {87}},\ \bibinfo
  {pages} {044028} (\bibinfo {year} {2013})},\ \Eprint
  {https://arxiv.org/abs/1211.2631} {arXiv:1211.2631 [gr-qc]} \BibitemShut
  {NoStop}%
\bibitem [{\citenamefont {Yang}\ \emph
  {et~al.}(2020{\natexlab{a}})\citenamefont {Yang}, \citenamefont {Wan},
  \citenamefont {Chen}, \citenamefont {Yang},\ and\ \citenamefont
  {Wang}}]{Yang:2020czk}%
  \BibitemOpen
  \bibfield  {author} {\bibinfo {author} {\bibfnamefont {S.-J.}\ \bibnamefont
  {Yang}}, \bibinfo {author} {\bibfnamefont {J.-J.}\ \bibnamefont {Wan}},
  \bibinfo {author} {\bibfnamefont {J.}~\bibnamefont {Chen}}, \bibinfo {author}
  {\bibfnamefont {J.}~\bibnamefont {Yang}},\ and\ \bibinfo {author}
  {\bibfnamefont {Y.-Q.}\ \bibnamefont {Wang}},\ }\bibfield  {title} {\bibinfo
  {title} {{Weak cosmic censorship conjecture for the novel $4D$ charged
  Einstein-Gauss-Bonnet black hole with test scalar field and particle}},\
  }\href {https://doi.org/10.1140/epjc/s10052-020-08511-9} {\bibfield
  {journal} {\bibinfo  {journal} {Eur. Phys. J. C}\ }\textbf {\bibinfo {volume}
  {80}},\ \bibinfo {pages} {937} (\bibinfo {year} {2020}{\natexlab{a}})},\
  \Eprint {https://arxiv.org/abs/2004.07934} {arXiv:2004.07934 [gr-qc]}
  \BibitemShut {NoStop}%
\bibitem [{\citenamefont {Barausse}\ \emph {et~al.}(2010)\citenamefont
  {Barausse}, \citenamefont {Cardoso},\ and\ \citenamefont {Khanna}}]{BaCK10}%
  \BibitemOpen
  \bibfield  {author} {\bibinfo {author} {\bibfnamefont {E.}~\bibnamefont
  {Barausse}}, \bibinfo {author} {\bibfnamefont {V.}~\bibnamefont {Cardoso}},\
  and\ \bibinfo {author} {\bibfnamefont {G.}~\bibnamefont {Khanna}},\
  }\bibfield  {title} {\bibinfo {title} {{Test bodies and naked singularities:
  Is the self-force the cosmic censor?}},\ }\href
  {https://doi.org/10.1103/PhysRevLett.105.261102} {\bibfield  {journal}
  {\bibinfo  {journal} {Phys. Rev. Lett.}\ }\textbf {\bibinfo {volume} {105}},\
  \bibinfo {pages} {261102} (\bibinfo {year} {2010})},\ \Eprint
  {https://arxiv.org/abs/1008.5159} {arXiv:1008.5159 [gr-qc]} \BibitemShut
  {NoStop}%
\bibitem [{\citenamefont {Barausse}\ \emph {et~al.}(2011)\citenamefont
  {Barausse}, \citenamefont {Cardoso},\ and\ \citenamefont {Khanna}}]{BaCK11}%
  \BibitemOpen
  \bibfield  {author} {\bibinfo {author} {\bibfnamefont {E.}~\bibnamefont
  {Barausse}}, \bibinfo {author} {\bibfnamefont {V.}~\bibnamefont {Cardoso}},\
  and\ \bibinfo {author} {\bibfnamefont {G.}~\bibnamefont {Khanna}},\
  }\bibfield  {title} {\bibinfo {title} {{Testing the Cosmic Censorship
  Conjecture with point particles: the effect of radiation reaction and the
  self-force}},\ }\href {https://doi.org/10.1103/PhysRevD.84.104006} {\bibfield
   {journal} {\bibinfo  {journal} {Phys. Rev. D}\ }\textbf {\bibinfo {volume}
  {84}},\ \bibinfo {pages} {104006} (\bibinfo {year} {2011})},\ \Eprint
  {https://arxiv.org/abs/1106.1692} {arXiv:1106.1692 [gr-qc]} \BibitemShut
  {NoStop}%
\bibitem [{\citenamefont {Zimmerman}\ \emph {et~al.}(2013)\citenamefont
  {Zimmerman}, \citenamefont {Vega}, \citenamefont {Poisson},\ and\
  \citenamefont {Haas}}]{ZVPH13}%
  \BibitemOpen
  \bibfield  {author} {\bibinfo {author} {\bibfnamefont {P.}~\bibnamefont
  {Zimmerman}}, \bibinfo {author} {\bibfnamefont {I.}~\bibnamefont {Vega}},
  \bibinfo {author} {\bibfnamefont {E.}~\bibnamefont {Poisson}},\ and\ \bibinfo
  {author} {\bibfnamefont {R.}~\bibnamefont {Haas}},\ }\bibfield  {title}
  {\bibinfo {title} {{Self-force as a cosmic censor}},\ }\href
  {https://doi.org/10.1103/PhysRevD.87.041501} {\bibfield  {journal} {\bibinfo
  {journal} {Phys. Rev. D}\ }\textbf {\bibinfo {volume} {87}},\ \bibinfo
  {pages} {041501} (\bibinfo {year} {2013})},\ \Eprint
  {https://arxiv.org/abs/1211.3889} {arXiv:1211.3889 [gr-qc]} \BibitemShut
  {NoStop}%
\bibitem [{\citenamefont {Colleoni}\ and\ \citenamefont
  {Barack}(2015)}]{CoBa15}%
  \BibitemOpen
  \bibfield  {author} {\bibinfo {author} {\bibfnamefont {M.}~\bibnamefont
  {Colleoni}}\ and\ \bibinfo {author} {\bibfnamefont {L.}~\bibnamefont
  {Barack}},\ }\bibfield  {title} {\bibinfo {title} {{Overspinning a Kerr black
  hole: the effect of self-force}},\ }\href
  {https://doi.org/10.1103/PhysRevD.91.104024} {\bibfield  {journal} {\bibinfo
  {journal} {Phys. Rev. D}\ }\textbf {\bibinfo {volume} {91}},\ \bibinfo
  {pages} {104024} (\bibinfo {year} {2015})},\ \Eprint
  {https://arxiv.org/abs/1501.07330} {arXiv:1501.07330 [gr-qc]} \BibitemShut
  {NoStop}%
\bibitem [{\citenamefont {Gwak}(2017)}]{Gwak17}%
  \BibitemOpen
  \bibfield  {author} {\bibinfo {author} {\bibfnamefont {B.}~\bibnamefont
  {Gwak}},\ }\bibfield  {title} {\bibinfo {title} {{Cosmic Censorship
  Conjecture in Kerr-Sen Black Hole}},\ }\href
  {https://doi.org/10.1103/PhysRevD.95.124050} {\bibfield  {journal} {\bibinfo
  {journal} {Phys. Rev. D}\ }\textbf {\bibinfo {volume} {95}},\ \bibinfo
  {pages} {124050} (\bibinfo {year} {2017})},\ \Eprint
  {https://arxiv.org/abs/1611.09640} {arXiv:1611.09640 [gr-qc]} \BibitemShut
  {NoStop}%
\bibitem [{\citenamefont {Liang}\ \emph {et~al.}(2019)\citenamefont {Liang},
  \citenamefont {Wei},\ and\ \citenamefont {Liu}}]{LiWL19}%
  \BibitemOpen
  \bibfield  {author} {\bibinfo {author} {\bibfnamefont {B.}~\bibnamefont
  {Liang}}, \bibinfo {author} {\bibfnamefont {S.-W.}\ \bibnamefont {Wei}},\
  and\ \bibinfo {author} {\bibfnamefont {Y.-X.}\ \bibnamefont {Liu}},\
  }\bibfield  {title} {\bibinfo {title} {{Weak cosmic censorship conjecture in
  Kerr black holes of modified gravity}},\ }\href
  {https://doi.org/10.1142/S0217732319500378} {\bibfield  {journal} {\bibinfo
  {journal} {Mod. Phys. Lett. A}\ }\textbf {\bibinfo {volume} {34}},\ \bibinfo
  {pages} {1950037} (\bibinfo {year} {2019})},\ \Eprint
  {https://arxiv.org/abs/1804.06966} {arXiv:1804.06966 [gr-qc]} \BibitemShut
  {NoStop}%
\bibitem [{\citenamefont {Sorce}\ and\ \citenamefont {Wald}(2017)}]{SoWa17}%
  \BibitemOpen
  \bibfield  {author} {\bibinfo {author} {\bibfnamefont {J.}~\bibnamefont
  {Sorce}}\ and\ \bibinfo {author} {\bibfnamefont {R.~M.}\ \bibnamefont
  {Wald}},\ }\bibfield  {title} {\bibinfo {title} {{Gedanken experiments to
  destroy a black hole. II. Kerr-Newman black holes cannot be overcharged or
  overspun}},\ }\href {https://doi.org/10.1103/PhysRevD.96.104014} {\bibfield
  {journal} {\bibinfo  {journal} {Phys. Rev. D}\ }\textbf {\bibinfo {volume}
  {96}},\ \bibinfo {pages} {104014} (\bibinfo {year} {2017})},\ \Eprint
  {https://arxiv.org/abs/1707.05862} {arXiv:1707.05862 [gr-qc]} \BibitemShut
  {NoStop}%
\bibitem [{\citenamefont {Sang}\ and\ \citenamefont {Jiang}(2021)}]{SaJi21}%
  \BibitemOpen
  \bibfield  {author} {\bibinfo {author} {\bibfnamefont {A.}~\bibnamefont
  {Sang}}\ and\ \bibinfo {author} {\bibfnamefont {J.}~\bibnamefont {Jiang}},\
  }\bibfield  {title} {\bibinfo {title} {{Gedanken experiments at high-order
  approximation: Kerr black hole cannot be overspun}},\ }\href
  {https://doi.org/10.1007/JHEP09(2021)095} {\bibfield  {journal} {\bibinfo
  {journal} {JHEP}\ }\textbf {\bibinfo {volume} {09}},\ \bibinfo {pages}
  {095}},\ \Eprint {https://arxiv.org/abs/2108.03454} {arXiv:2108.03454
  [gr-qc]} \BibitemShut {NoStop}%
\bibitem [{\citenamefont {Zhang}\ and\ \citenamefont {Jiang}(2020)}]{ZhJi20}%
  \BibitemOpen
  \bibfield  {author} {\bibinfo {author} {\bibfnamefont {M.}~\bibnamefont
  {Zhang}}\ and\ \bibinfo {author} {\bibfnamefont {J.}~\bibnamefont {Jiang}},\
  }\bibfield  {title} {\bibinfo {title} {{New gedanken experiment on
  higher-dimensional asymptotically AdS Reissner-Nordstr\"om black hole}},\
  }\href {https://doi.org/10.1140/epjc/s10052-020-08475-w} {\bibfield
  {journal} {\bibinfo  {journal} {Eur. Phys. J. C}\ }\textbf {\bibinfo {volume}
  {80}},\ \bibinfo {pages} {890} (\bibinfo {year} {2020})},\ \Eprint
  {https://arxiv.org/abs/2009.07681} {arXiv:2009.07681 [gr-qc]} \BibitemShut
  {NoStop}%
\bibitem [{\citenamefont {Qu}\ \emph {et~al.}(2020)\citenamefont {Qu},
  \citenamefont {Yang}, \citenamefont {Wang},\ and\ \citenamefont
  {Ren}}]{QYWR20}%
  \BibitemOpen
  \bibfield  {author} {\bibinfo {author} {\bibfnamefont {F.}~\bibnamefont
  {Qu}}, \bibinfo {author} {\bibfnamefont {S.-J.}\ \bibnamefont {Yang}},
  \bibinfo {author} {\bibfnamefont {Z.}~\bibnamefont {Wang}},\ and\ \bibinfo
  {author} {\bibfnamefont {J.-R.}\ \bibnamefont {Ren}},\ }\bibfield  {title}
  {\bibinfo {title} {{Weak cosmic censorship conjecture is not violated for a
  rotating linear dilaton black hole}},\ }\href@noop {} {\  (\bibinfo {year}
  {2020})},\ \Eprint {https://arxiv.org/abs/2008.09950} {arXiv:2008.09950
  [gr-qc]} \BibitemShut {NoStop}%
\bibitem [{\citenamefont {Wang}\ and\ \citenamefont {Jiang}(2020)}]{WaJi20}%
  \BibitemOpen
  \bibfield  {author} {\bibinfo {author} {\bibfnamefont {X.-Y.}\ \bibnamefont
  {Wang}}\ and\ \bibinfo {author} {\bibfnamefont {J.}~\bibnamefont {Jiang}},\
  }\bibfield  {title} {\bibinfo {title} {{Gedanken experiments at high-order
  approximation: nearly extremal Reissner-Nordstr\"om black holes cannot be
  overcharged}},\ }\href {https://doi.org/10.1007/JHEP05(2020)161} {\bibfield
  {journal} {\bibinfo  {journal} {JHEP}\ }\textbf {\bibinfo {volume} {05}},\
  \bibinfo {pages} {161}},\ \Eprint {https://arxiv.org/abs/2004.12120}
  {arXiv:2004.12120 [hep-th]} \BibitemShut {NoStop}%
\bibitem [{\citenamefont {Jiang}\ and\ \citenamefont {Gao}(2020)}]{JiGa20}%
  \BibitemOpen
  \bibfield  {author} {\bibinfo {author} {\bibfnamefont {J.}~\bibnamefont
  {Jiang}}\ and\ \bibinfo {author} {\bibfnamefont {Y.}~\bibnamefont {Gao}},\
  }\bibfield  {title} {\bibinfo {title} {{Investigating the gedanken experiment
  to destroy the event horizon of a regular black hole}},\ }\href
  {https://doi.org/10.1103/PhysRevD.101.084005} {\bibfield  {journal} {\bibinfo
   {journal} {Phys. Rev. D}\ }\textbf {\bibinfo {volume} {101}},\ \bibinfo
  {pages} {084005} (\bibinfo {year} {2020})},\ \Eprint
  {https://arxiv.org/abs/2003.07501} {arXiv:2003.07501 [hep-th]} \BibitemShut
  {NoStop}%
\bibitem [{\citenamefont {Chen}\ \emph {et~al.}(2019)\citenamefont {Chen},
  \citenamefont {Lin},\ and\ \citenamefont {Ning}}]{ChLN19}%
  \BibitemOpen
  \bibfield  {author} {\bibinfo {author} {\bibfnamefont {B.}~\bibnamefont
  {Chen}}, \bibinfo {author} {\bibfnamefont {F.-L.}\ \bibnamefont {Lin}},\ and\
  \bibinfo {author} {\bibfnamefont {B.}~\bibnamefont {Ning}},\ }\bibfield
  {title} {\bibinfo {title} {{Gedanken Experiments to Destroy a BTZ Black
  Hole}},\ }\href {https://doi.org/10.1103/PhysRevD.100.044043} {\bibfield
  {journal} {\bibinfo  {journal} {Phys. Rev. D}\ }\textbf {\bibinfo {volume}
  {100}},\ \bibinfo {pages} {044043} (\bibinfo {year} {2019})},\ \Eprint
  {https://arxiv.org/abs/1902.00949} {arXiv:1902.00949 [gr-qc]} \BibitemShut
  {NoStop}%
\bibitem [{\citenamefont {Shaymatov}\ and\ \citenamefont
  {Dadhich}(2022)}]{Shaymatov:2020byu}%
  \BibitemOpen
  \bibfield  {author} {\bibinfo {author} {\bibfnamefont {S.}~\bibnamefont
  {Shaymatov}}\ and\ \bibinfo {author} {\bibfnamefont {N.}~\bibnamefont
  {Dadhich}},\ }\bibfield  {title} {\bibinfo {title} {{Weak cosmic censorship
  conjecture in the pure Lovelock gravity}},\ }\href
  {https://doi.org/10.1088/1475-7516/2022/10/060} {\bibfield  {journal}
  {\bibinfo  {journal} {JCAP}\ }\textbf {\bibinfo {volume} {10}},\ \bibinfo
  {pages} {060}},\ \Eprint {https://arxiv.org/abs/2008.04092} {arXiv:2008.04092
  [gr-qc]} \BibitemShut {NoStop}%
\bibitem [{\citenamefont {D\"uzta\c{s}}(2019)}]{Duztas:2019ick}%
  \BibitemOpen
  \bibfield  {author} {\bibinfo {author} {\bibfnamefont {K.}~\bibnamefont
  {D\"uzta\c{s}}},\ }\bibfield  {title} {\bibinfo {title}
  {{Kerr\textendash{}Newman black holes can be generically overspun}},\ }\href
  {https://doi.org/10.1140/epjc/s10052-019-6851-z} {\bibfield  {journal}
  {\bibinfo  {journal} {Eur. Phys. J. C}\ }\textbf {\bibinfo {volume} {79}},\
  \bibinfo {pages} {316} (\bibinfo {year} {2019})},\ \Eprint
  {https://arxiv.org/abs/1904.05185} {arXiv:1904.05185 [gr-qc]} \BibitemShut
  {NoStop}%
\bibitem [{\citenamefont {Gwak}(2021)}]{Gwak:2021tcl}%
  \BibitemOpen
  \bibfield  {author} {\bibinfo {author} {\bibfnamefont {B.}~\bibnamefont
  {Gwak}},\ }\bibfield  {title} {\bibinfo {title} {{Weak cosmic censorship
  conjecture in Kerr-Newman-(anti-)de Sitter black hole with charged scalar
  field}},\ }\href {https://doi.org/10.1088/1475-7516/2021/10/012} {\bibfield
  {journal} {\bibinfo  {journal} {JCAP}\ }\textbf {\bibinfo {volume} {10}},\
  \bibinfo {pages} {012}},\ \Eprint {https://arxiv.org/abs/2105.07226}
  {arXiv:2105.07226 [gr-qc]} \BibitemShut {NoStop}%
\bibitem [{\citenamefont {Natario}\ \emph {et~al.}(2016)\citenamefont
  {Natario}, \citenamefont {Queimada},\ and\ \citenamefont {Vicente}}]{NaQV16}%
  \BibitemOpen
  \bibfield  {author} {\bibinfo {author} {\bibfnamefont {J.}~\bibnamefont
  {Natario}}, \bibinfo {author} {\bibfnamefont {L.}~\bibnamefont {Queimada}},\
  and\ \bibinfo {author} {\bibfnamefont {R.}~\bibnamefont {Vicente}},\
  }\bibfield  {title} {\bibinfo {title} {{Test fields cannot destroy extremal
  black holes}},\ }\href {https://doi.org/10.1088/0264-9381/33/17/175002}
  {\bibfield  {journal} {\bibinfo  {journal} {Class. Quant. Grav.}\ }\textbf
  {\bibinfo {volume} {33}},\ \bibinfo {pages} {175002} (\bibinfo {year}
  {2016})},\ \Eprint {https://arxiv.org/abs/1601.06809} {arXiv:1601.06809
  [gr-qc]} \BibitemShut {NoStop}%
\bibitem [{\citenamefont {Yoshida}\ and\ \citenamefont
  {Yoshimura}(2024)}]{Yoshida:2024txh}%
  \BibitemOpen
  \bibfield  {author} {\bibinfo {author} {\bibfnamefont {D.}~\bibnamefont
  {Yoshida}}\ and\ \bibinfo {author} {\bibfnamefont {K.}~\bibnamefont
  {Yoshimura}},\ }\bibfield  {title} {\bibinfo {title} {{The First Law and Weak
  Cosmic Censorship for de Sitter Black Holes}},\ }\href@noop {} {\  (\bibinfo
  {year} {2024})},\ \Eprint {https://arxiv.org/abs/2411.19535}
  {arXiv:2411.19535 [gr-qc]} \BibitemShut {NoStop}%
\bibitem [{\citenamefont {Yang}\ \emph {et~al.}(2023)\citenamefont {Yang},
  \citenamefont {Guo}, \citenamefont {Wei},\ and\ \citenamefont
  {Liu}}]{Yang:2023hll}%
  \BibitemOpen
  \bibfield  {author} {\bibinfo {author} {\bibfnamefont {S.-J.}\ \bibnamefont
  {Yang}}, \bibinfo {author} {\bibfnamefont {W.-D.}\ \bibnamefont {Guo}},
  \bibinfo {author} {\bibfnamefont {S.-W.}\ \bibnamefont {Wei}},\ and\ \bibinfo
  {author} {\bibfnamefont {Y.-X.}\ \bibnamefont {Liu}},\ }\bibfield  {title}
  {\bibinfo {title} {{First law of black hole thermodynamics and the weak
  cosmic censorship conjecture for Kerr\textendash{}Newman Taub\textendash{}NUT
  black holes}},\ }\href {https://doi.org/10.1140/epjc/s10052-023-12265-5}
  {\bibfield  {journal} {\bibinfo  {journal} {Eur. Phys. J. C}\ }\textbf
  {\bibinfo {volume} {83}},\ \bibinfo {pages} {1111} (\bibinfo {year}
  {2023})},\ \Eprint {https://arxiv.org/abs/2306.05266} {arXiv:2306.05266
  [gr-qc]} \BibitemShut {NoStop}%
\bibitem [{\citenamefont {Yang}\ \emph
  {et~al.}(2020{\natexlab{b}})\citenamefont {Yang}, \citenamefont {Chen},
  \citenamefont {Wan}, \citenamefont {Wei},\ and\ \citenamefont
  {Liu}}]{Yang:2020iat}%
  \BibitemOpen
  \bibfield  {author} {\bibinfo {author} {\bibfnamefont {S.-J.}\ \bibnamefont
  {Yang}}, \bibinfo {author} {\bibfnamefont {J.}~\bibnamefont {Chen}}, \bibinfo
  {author} {\bibfnamefont {J.-J.}\ \bibnamefont {Wan}}, \bibinfo {author}
  {\bibfnamefont {S.-W.}\ \bibnamefont {Wei}},\ and\ \bibinfo {author}
  {\bibfnamefont {Y.-X.}\ \bibnamefont {Liu}},\ }\bibfield  {title} {\bibinfo
  {title} {{Weak cosmic censorship conjecture for a Kerr-Taub-NUT black hole
  with a test scalar field and particle}},\ }\href
  {https://doi.org/10.1103/PhysRevD.101.064048} {\bibfield  {journal} {\bibinfo
   {journal} {Phys. Rev. D}\ }\textbf {\bibinfo {volume} {101}},\ \bibinfo
  {pages} {064048} (\bibinfo {year} {2020}{\natexlab{b}})},\ \Eprint
  {https://arxiv.org/abs/2001.03106} {arXiv:2001.03106 [gr-qc]} \BibitemShut
  {NoStop}%
\bibitem [{\citenamefont {Feng}\ \emph {et~al.}(2021)\citenamefont {Feng},
  \citenamefont {Yang}, \citenamefont {Tan}, \citenamefont {Yang},\ and\
  \citenamefont {Liu}}]{Feng:2020tyc}%
  \BibitemOpen
  \bibfield  {author} {\bibinfo {author} {\bibfnamefont {W.-B.}\ \bibnamefont
  {Feng}}, \bibinfo {author} {\bibfnamefont {S.-J.}\ \bibnamefont {Yang}},
  \bibinfo {author} {\bibfnamefont {Q.}~\bibnamefont {Tan}}, \bibinfo {author}
  {\bibfnamefont {J.}~\bibnamefont {Yang}},\ and\ \bibinfo {author}
  {\bibfnamefont {Y.-X.}\ \bibnamefont {Liu}},\ }\bibfield  {title} {\bibinfo
  {title} {{Overcharging a Reissner-Nordstr\"om Taub-NUT regular black hole}},\
  }\href {https://doi.org/10.1007/s11433-020-1659-0} {\bibfield  {journal}
  {\bibinfo  {journal} {Sci. China Phys. Mech. Astron.}\ }\textbf {\bibinfo
  {volume} {64}},\ \bibinfo {pages} {260411} (\bibinfo {year} {2021})},\
  \Eprint {https://arxiv.org/abs/2009.12846} {arXiv:2009.12846 [gr-qc]}
  \BibitemShut {NoStop}%
\bibitem [{\citenamefont {Gao}\ and\ \citenamefont {Gao}(2022)}]{Gao:2022ckf}%
  \BibitemOpen
  \bibfield  {author} {\bibinfo {author} {\bibfnamefont {Y.}~\bibnamefont
  {Gao}}\ and\ \bibinfo {author} {\bibfnamefont {S.}~\bibnamefont {Gao}},\
  }\bibfield  {title} {\bibinfo {title} {{Testing the weak cosmic censorship
  conjecture for extremal magnetized Kerr\textendash{}Newman black holes}},\
  }\href {https://doi.org/10.1140/epjc/s10052-022-10709-y} {\bibfield
  {journal} {\bibinfo  {journal} {Eur. Phys. J. C}\ }\textbf {\bibinfo {volume}
  {82}},\ \bibinfo {pages} {763} (\bibinfo {year} {2022})},\ \Eprint
  {https://arxiv.org/abs/2208.00703} {arXiv:2208.00703 [gr-qc]} \BibitemShut
  {NoStop}%
\bibitem [{\citenamefont {Lin}\ and\ \citenamefont {Ning}(2024)}]{Lin:2024deg}%
  \BibitemOpen
  \bibfield  {author} {\bibinfo {author} {\bibfnamefont {F.-L.}\ \bibnamefont
  {Lin}}\ and\ \bibinfo {author} {\bibfnamefont {B.}~\bibnamefont {Ning}},\
  }\bibfield  {title} {\bibinfo {title} {{Violation of weak cosmic censorship
  in de Sitter space}},\ }\href {https://doi.org/10.1103/PhysRevD.110.044057}
  {\bibfield  {journal} {\bibinfo  {journal} {Phys. Rev. D}\ }\textbf {\bibinfo
  {volume} {110}},\ \bibinfo {pages} {044057} (\bibinfo {year} {2024})},\
  \Eprint {https://arxiv.org/abs/2405.07728} {arXiv:2405.07728 [hep-th]}
  \BibitemShut {NoStop}%
\bibitem [{\citenamefont {Ali}\ \emph {et~al.}(2023)\citenamefont {Ali},
  \citenamefont {El~Moumni}, \citenamefont {Khalloufi},\ and\ \citenamefont
  {Masmar}}]{Ali:2023iuz}%
  \BibitemOpen
  \bibfield  {author} {\bibinfo {author} {\bibfnamefont {M.~S.}\ \bibnamefont
  {Ali}}, \bibinfo {author} {\bibfnamefont {H.}~\bibnamefont {El~Moumni}},
  \bibinfo {author} {\bibfnamefont {J.}~\bibnamefont {Khalloufi}},\ and\
  \bibinfo {author} {\bibfnamefont {K.}~\bibnamefont {Masmar}},\ }\bibfield
  {title} {\bibinfo {title} {{Revisiting the second law and weak cosmic
  censorship conjecture in high-dimensional charged-AdS black hole: an
  additional assumption}},\ }\href {https://doi.org/10.1007/JHEP03(2023)160}
  {\bibfield  {journal} {\bibinfo  {journal} {JHEP}\ }\textbf {\bibinfo
  {volume} {03}},\ \bibinfo {pages} {160}},\ \Eprint
  {https://arxiv.org/abs/2302.07026} {arXiv:2302.07026 [hep-th]} \BibitemShut
  {NoStop}%
\bibitem [{\citenamefont {Bekenstein}(1973{\natexlab{b}})}]{Bekenstein:1973mi}%
  \BibitemOpen
  \bibfield  {author} {\bibinfo {author} {\bibfnamefont {J.~D.}\ \bibnamefont
  {Bekenstein}},\ }\bibfield  {title} {\bibinfo {title} {{Extraction of energy
  and charge from a black hole}},\ }\href
  {https://doi.org/10.1103/PhysRevD.7.949} {\bibfield  {journal} {\bibinfo
  {journal} {Phys. Rev. D}\ }\textbf {\bibinfo {volume} {7}},\ \bibinfo {pages}
  {949} (\bibinfo {year} {1973}{\natexlab{b}})}\BibitemShut {NoStop}%
\bibitem [{\citenamefont {Lin}\ \emph {et~al.}(2023)\citenamefont {Lin},
  \citenamefont {Ning},\ and\ \citenamefont {Chen}}]{Lin:2022ndf}%
  \BibitemOpen
  \bibfield  {author} {\bibinfo {author} {\bibfnamefont {F.-L.}\ \bibnamefont
  {Lin}}, \bibinfo {author} {\bibfnamefont {B.}~\bibnamefont {Ning}},\ and\
  \bibinfo {author} {\bibfnamefont {Y.}~\bibnamefont {Chen}},\ }\bibfield
  {title} {\bibinfo {title} {{Weak cosmic censorship and the second law of
  black hole thermodynamics in higher derivative gravity}},\ }\href
  {https://doi.org/10.1103/PhysRevD.108.044025} {\bibfield  {journal} {\bibinfo
   {journal} {Phys. Rev. D}\ }\textbf {\bibinfo {volume} {108}},\ \bibinfo
  {pages} {044025} (\bibinfo {year} {2023})},\ \Eprint
  {https://arxiv.org/abs/2211.17225} {arXiv:2211.17225 [hep-th]} \BibitemShut
  {NoStop}%
\bibitem [{\citenamefont {Gwak}(2018)}]{Gwak18}%
  \BibitemOpen
  \bibfield  {author} {\bibinfo {author} {\bibfnamefont {B.}~\bibnamefont
  {Gwak}},\ }\bibfield  {title} {\bibinfo {title} {{Weak Cosmic Censorship
  Conjecture in Kerr-(Anti-)de Sitter Black Hole with Scalar Field}},\ }\href
  {https://doi.org/10.1007/JHEP09(2018)081} {\bibfield  {journal} {\bibinfo
  {journal} {JHEP}\ }\textbf {\bibinfo {volume} {09}},\ \bibinfo {pages}
  {081}},\ \Eprint {https://arxiv.org/abs/1807.10630} {arXiv:1807.10630
  [gr-qc]} \BibitemShut {NoStop}%
\bibitem [{\citenamefont {Page}(1976)}]{Page:1976jj}%
  \BibitemOpen
  \bibfield  {author} {\bibinfo {author} {\bibfnamefont {D.~N.}\ \bibnamefont
  {Page}},\ }\bibfield  {title} {\bibinfo {title} {{Dirac Equation Around a
  Charged, Rotating Black Hole}},\ }\href
  {https://doi.org/10.1103/PhysRevD.14.1509} {\bibfield  {journal} {\bibinfo
  {journal} {Phys. Rev. D}\ }\textbf {\bibinfo {volume} {14}},\ \bibinfo
  {pages} {1509} (\bibinfo {year} {1976})}\BibitemShut {NoStop}%
\bibitem [{\citenamefont {Lee}(1977)}]{Lee:1977gk}%
  \BibitemOpen
  \bibfield  {author} {\bibinfo {author} {\bibfnamefont {C.~H.}\ \bibnamefont
  {Lee}},\ }\bibfield  {title} {\bibinfo {title} {{Massive Spin 1/2 Wave Around
  a Kerr-Newman Black Hole}},\ }\href
  {https://doi.org/10.1016/0370-2693(77)90189-7} {\bibfield  {journal}
  {\bibinfo  {journal} {Phys. Lett. B}\ }\textbf {\bibinfo {volume} {68}},\
  \bibinfo {pages} {152} (\bibinfo {year} {1977})}\BibitemShut {NoStop}%
\bibitem [{\citenamefont {Bjorken}\ and\ \citenamefont
  {Drell}(1965)}]{Bjorken:1979dk}%
  \BibitemOpen
  \bibfield  {author} {\bibinfo {author} {\bibfnamefont {J.~D.}\ \bibnamefont
  {Bjorken}}\ and\ \bibinfo {author} {\bibfnamefont {S.~D.}\ \bibnamefont
  {Drell}},\ }\href@noop {} {\emph {\bibinfo {title} {{Relativistic Quantum
  Field Theory. (McGraw-Hill, New York)}}}}\ (\bibinfo {year}
  {1965})\BibitemShut {NoStop}%
\bibitem [{\citenamefont {Unruh}(1973)}]{Unruh:1973bda}%
  \BibitemOpen
  \bibfield  {author} {\bibinfo {author} {\bibfnamefont {W.}~\bibnamefont
  {Unruh}},\ }\bibfield  {title} {\bibinfo {title} {{Separability of the
  Neutrino Equations in a Kerr Background}},\ }\href
  {https://doi.org/10.1103/PhysRevLett.31.1265} {\bibfield  {journal} {\bibinfo
   {journal} {Phys. Rev. Lett.}\ }\textbf {\bibinfo {volume} {31}},\ \bibinfo
  {pages} {1265} (\bibinfo {year} {1973})}\BibitemShut {NoStop}%
\bibitem [{\citenamefont {Brill}\ and\ \citenamefont
  {Wheeler}(1957)}]{Brill:1957fx}%
  \BibitemOpen
  \bibfield  {author} {\bibinfo {author} {\bibfnamefont {D.~R.}\ \bibnamefont
  {Brill}}\ and\ \bibinfo {author} {\bibfnamefont {J.~A.}\ \bibnamefont
  {Wheeler}},\ }\bibfield  {title} {\bibinfo {title} {{Interaction of neutrinos
  and gravitational fields}},\ }\href
  {https://doi.org/10.1103/RevModPhys.29.465} {\bibfield  {journal} {\bibinfo
  {journal} {Rev. Mod. Phys.}\ }\textbf {\bibinfo {volume} {29}},\ \bibinfo
  {pages} {465} (\bibinfo {year} {1957})}\BibitemShut {NoStop}%
\bibitem [{\citenamefont {Hawking}(1971)}]{Hawking:1971tu}%
  \BibitemOpen
  \bibfield  {author} {\bibinfo {author} {\bibfnamefont {S.~W.}\ \bibnamefont
  {Hawking}},\ }\bibfield  {title} {\bibinfo {title} {{Gravitational radiation
  from colliding black holes}},\ }\href
  {https://doi.org/10.1103/PhysRevLett.26.1344} {\bibfield  {journal} {\bibinfo
   {journal} {Phys. Rev. Lett.}\ }\textbf {\bibinfo {volume} {26}},\ \bibinfo
  {pages} {1344} (\bibinfo {year} {1971})}\BibitemShut {NoStop}%
\bibitem [{\citenamefont {Hennigar}\ \emph {et~al.}(2019)\citenamefont
  {Hennigar}, \citenamefont {Kubiz\v{n}\'ak},\ and\ \citenamefont
  {Mann}}]{Hennigar:2019ive}%
  \BibitemOpen
  \bibfield  {author} {\bibinfo {author} {\bibfnamefont {R.~A.}\ \bibnamefont
  {Hennigar}}, \bibinfo {author} {\bibfnamefont {D.}~\bibnamefont
  {Kubiz\v{n}\'ak}},\ and\ \bibinfo {author} {\bibfnamefont {R.~B.}\
  \bibnamefont {Mann}},\ }\bibfield  {title} {\bibinfo {title} {{Thermodynamics
  of Lorentzian Taub-NUT spacetimes}},\ }\href
  {https://doi.org/10.1103/PhysRevD.100.064055} {\bibfield  {journal} {\bibinfo
   {journal} {Phys. Rev. D}\ }\textbf {\bibinfo {volume} {100}},\ \bibinfo
  {pages} {064055} (\bibinfo {year} {2019})},\ \Eprint
  {https://arxiv.org/abs/1903.08668} {arXiv:1903.08668 [hep-th]} \BibitemShut
  {NoStop}%
\bibitem [{\citenamefont {Bordo}\ \emph {et~al.}(2019)\citenamefont {Bordo},
  \citenamefont {Gray}, \citenamefont {Hennigar},\ and\ \citenamefont
  {Kubiz\v{n}\'ak}}]{Bordo:2019tyh}%
  \BibitemOpen
  \bibfield  {author} {\bibinfo {author} {\bibfnamefont {A.~B.}\ \bibnamefont
  {Bordo}}, \bibinfo {author} {\bibfnamefont {F.}~\bibnamefont {Gray}},
  \bibinfo {author} {\bibfnamefont {R.~A.}\ \bibnamefont {Hennigar}},\ and\
  \bibinfo {author} {\bibfnamefont {D.}~\bibnamefont {Kubiz\v{n}\'ak}},\
  }\bibfield  {title} {\bibinfo {title} {{Misner Gravitational Charges and
  Variable String Strengths}},\ }\href
  {https://doi.org/10.1088/1361-6382/ab3d4d} {\bibfield  {journal} {\bibinfo
  {journal} {Class. Quant. Grav.}\ }\textbf {\bibinfo {volume} {36}},\ \bibinfo
  {pages} {194001} (\bibinfo {year} {2019})},\ \Eprint
  {https://arxiv.org/abs/1905.03785} {arXiv:1905.03785 [hep-th]} \BibitemShut
  {NoStop}%
\bibitem [{\citenamefont {Chen}\ and\ \citenamefont
  {Jiang}(2019)}]{Chen:2019uhp}%
  \BibitemOpen
  \bibfield  {author} {\bibinfo {author} {\bibfnamefont {Z.}~\bibnamefont
  {Chen}}\ and\ \bibinfo {author} {\bibfnamefont {J.}~\bibnamefont {Jiang}},\
  }\bibfield  {title} {\bibinfo {title} {{General Smarr relation and first law
  of a NUT dyonic black hole}},\ }\href
  {https://doi.org/10.1103/PhysRevD.100.104016} {\bibfield  {journal} {\bibinfo
   {journal} {Phys. Rev. D}\ }\textbf {\bibinfo {volume} {100}},\ \bibinfo
  {pages} {104016} (\bibinfo {year} {2019})},\ \Eprint
  {https://arxiv.org/abs/1910.10107} {arXiv:1910.10107 [hep-th]} \BibitemShut
  {NoStop}%
\bibitem [{\citenamefont {Wu}\ and\ \citenamefont {Wu}(2019)}]{Wu:2019pzr}%
  \BibitemOpen
  \bibfield  {author} {\bibinfo {author} {\bibfnamefont {S.-Q.}\ \bibnamefont
  {Wu}}\ and\ \bibinfo {author} {\bibfnamefont {D.}~\bibnamefont {Wu}},\
  }\bibfield  {title} {\bibinfo {title} {{Thermodynamical hairs of the
  four-dimensional Taub-Newman-Unti-Tamburino spacetimes}},\ }\href
  {https://doi.org/10.1103/PhysRevD.100.101501} {\bibfield  {journal} {\bibinfo
   {journal} {Phys. Rev. D}\ }\textbf {\bibinfo {volume} {100}},\ \bibinfo
  {pages} {101501} (\bibinfo {year} {2019})},\ \Eprint
  {https://arxiv.org/abs/1909.07776} {arXiv:1909.07776 [hep-th]} \BibitemShut
  {NoStop}%
\bibitem [{\citenamefont {Liu}\ \emph {et~al.}(2022)\citenamefont {Liu},
  \citenamefont {Lu},\ and\ \citenamefont {Ma}}]{Liu:2022wku}%
  \BibitemOpen
  \bibfield  {author} {\bibinfo {author} {\bibfnamefont {H.-S.}\ \bibnamefont
  {Liu}}, \bibinfo {author} {\bibfnamefont {H.}~\bibnamefont {Lu}},\ and\
  \bibinfo {author} {\bibfnamefont {L.}~\bibnamefont {Ma}},\ }\bibfield
  {title} {\bibinfo {title} {{Thermodynamics of Taub-NUT and Plebanski
  solutions}},\ }\href {https://doi.org/10.1007/JHEP10(2022)174} {\bibfield
  {journal} {\bibinfo  {journal} {JHEP}\ }\textbf {\bibinfo {volume} {10}},\
  \bibinfo {pages} {174}},\ \Eprint {https://arxiv.org/abs/2208.05494}
  {arXiv:2208.05494 [gr-qc]} \BibitemShut {NoStop}%
\bibitem [{\citenamefont {D\"uzta\c{s}}(2018{\natexlab{a}})}]{Duztas:2017lxk}%
  \BibitemOpen
  \bibfield  {author} {\bibinfo {author} {\bibfnamefont {K.}~\bibnamefont
  {D\"uzta\c{s}}},\ }\bibfield  {title} {\bibinfo {title} {{Can test fields
  destroy the event horizon in the Kerr\textendash{}Taub\textendash{}NUT
  spacetime?}},\ }\href {https://doi.org/10.1088/1361-6382/aaa4e0} {\bibfield
  {journal} {\bibinfo  {journal} {Class. Quant. Grav.}\ }\textbf {\bibinfo
  {volume} {35}},\ \bibinfo {pages} {045008} (\bibinfo {year}
  {2018}{\natexlab{a}})},\ \Eprint {https://arxiv.org/abs/1710.06610}
  {arXiv:1710.06610 [gr-qc]} \BibitemShut {NoStop}%
\bibitem [{\citenamefont {D\"uzta\c{s}}(2016)}]{Duztas:2016xfg}%
  \BibitemOpen
  \bibfield  {author} {\bibinfo {author} {\bibfnamefont {K.}~\bibnamefont
  {D\"uzta\c{s}}},\ }\bibfield  {title} {\bibinfo {title} {{Overspinning BTZ
  black holes with test particles and fields}},\ }\href
  {https://doi.org/10.1103/PhysRevD.94.124031} {\bibfield  {journal} {\bibinfo
  {journal} {Phys. Rev. D}\ }\textbf {\bibinfo {volume} {94}},\ \bibinfo
  {pages} {124031} (\bibinfo {year} {2016})},\ \Eprint
  {https://arxiv.org/abs/1701.07241} {arXiv:1701.07241 [gr-qc]} \BibitemShut
  {NoStop}%
\bibitem [{\citenamefont {D\"uzta\c{s}}(2018{\natexlab{b}})}]{Duztas:2018adf}%
  \BibitemOpen
  \bibfield  {author} {\bibinfo {author} {\bibfnamefont {K.}~\bibnamefont
  {D\"uzta\c{s}}},\ }\bibfield  {title} {\bibinfo {title} {{Over-spinning
  Kerr\textendash{}Sen black holes with test fields}},\ }\href
  {https://doi.org/10.1142/S0218271819500445} {\bibfield  {journal} {\bibinfo
  {journal} {Int. J. Mod. Phys. D}\ }\textbf {\bibinfo {volume} {28}},\
  \bibinfo {pages} {1950044} (\bibinfo {year} {2018}{\natexlab{b}})},\ \Eprint
  {https://arxiv.org/abs/1811.03452} {arXiv:1811.03452 [gr-qc]} \BibitemShut
  {NoStop}%
\bibitem [{\citenamefont {Chen}(2020)}]{Chen:2018yah}%
  \BibitemOpen
  \bibfield  {author} {\bibinfo {author} {\bibfnamefont {D.}~\bibnamefont
  {Chen}},\ }\bibfield  {title} {\bibinfo {title} {{Weak cosmic censorship
  conjecture in BTZ black holes with scalar fields}},\ }\href
  {https://doi.org/10.1088/1674-1137/44/1/015101} {\bibfield  {journal}
  {\bibinfo  {journal} {Chin. Phys. C}\ }\textbf {\bibinfo {volume} {44}},\
  \bibinfo {pages} {015101} (\bibinfo {year} {2020})},\ \Eprint
  {https://arxiv.org/abs/1812.03459} {arXiv:1812.03459 [gr-qc]} \BibitemShut
  {NoStop}%
\bibitem [{\citenamefont {Gwak}(2020)}]{Gwak:2019rcz}%
  \BibitemOpen
  \bibfield  {author} {\bibinfo {author} {\bibfnamefont {B.}~\bibnamefont
  {Gwak}},\ }\bibfield  {title} {\bibinfo {title} {{Weak Cosmic Censorship in
  Kerr-Sen Black Hole under Charged Scalar Field}},\ }\href
  {https://doi.org/10.1088/1475-7516/2020/03/058} {\bibfield  {journal}
  {\bibinfo  {journal} {JCAP}\ }\textbf {\bibinfo {volume} {03}},\ \bibinfo
  {pages} {058}},\ \Eprint {https://arxiv.org/abs/1910.13329} {arXiv:1910.13329
  [gr-qc]} \BibitemShut {NoStop}%
\bibitem [{\citenamefont {T\'oth}(2016)}]{Toth:2015cda}%
  \BibitemOpen
  \bibfield  {author} {\bibinfo {author} {\bibfnamefont {G.~Z.}\ \bibnamefont
  {T\'oth}},\ }\bibfield  {title} {\bibinfo {title} {{Weak cosmic censorship,
  dyonic Kerr\textendash{}Newman black holes and Dirac fields}},\ }\href
  {https://doi.org/10.1088/0264-9381/33/11/115012} {\bibfield  {journal}
  {\bibinfo  {journal} {Class. Quant. Grav.}\ }\textbf {\bibinfo {volume}
  {33}},\ \bibinfo {pages} {115012} (\bibinfo {year} {2016})},\ \Eprint
  {https://arxiv.org/abs/1509.02878} {arXiv:1509.02878 [gr-qc]} \BibitemShut
  {NoStop}%
\bibitem [{\citenamefont {Toth}(2012)}]{Toth:2012vvy}%
  \BibitemOpen
  \bibfield  {author} {\bibinfo {author} {\bibfnamefont {G.~Z.}\ \bibnamefont
  {Toth}},\ }\bibfield  {title} {\bibinfo {title} {{Test of the weak cosmic
  censorship conjecture with a charged scalar field and dyonic Kerr-Newman
  black holes}},\ }\href {https://doi.org/10.1007/s10714-012-1374-z} {\bibfield
   {journal} {\bibinfo  {journal} {Gen. Rel. Grav.}\ }\textbf {\bibinfo
  {volume} {44}},\ \bibinfo {pages} {2019} (\bibinfo {year} {2012})},\ \Eprint
  {https://arxiv.org/abs/1112.2382} {arXiv:1112.2382 [gr-qc]} \BibitemShut
  {NoStop}%
\bibitem [{\citenamefont {Wu}\ \emph {et~al.}(2024)\citenamefont {Wu},
  \citenamefont {Khodabakhshi},\ and\ \citenamefont {Lu}}]{Wu:2024ucf}%
  \BibitemOpen
  \bibfield  {author} {\bibinfo {author} {\bibfnamefont {P.-Y.}\ \bibnamefont
  {Wu}}, \bibinfo {author} {\bibfnamefont {H.}~\bibnamefont {Khodabakhshi}},\
  and\ \bibinfo {author} {\bibfnamefont {H.}~\bibnamefont {Lu}},\ }\bibfield
  {title} {\bibinfo {title} {{Weak cosmic censorship conjecture cannot be
  violated in gedanken experiments}},\ }\href
  {https://doi.org/10.1103/PhysRevD.110.104019} {\bibfield  {journal} {\bibinfo
   {journal} {Phys. Rev. D}\ }\textbf {\bibinfo {volume} {110}},\ \bibinfo
  {pages} {104019} (\bibinfo {year} {2024})},\ \Eprint
  {https://arxiv.org/abs/2408.09444} {arXiv:2408.09444 [gr-qc]} \BibitemShut
  {NoStop}%
\bibitem [{\citenamefont {Lu}\ \emph {et~al.}(2025)\citenamefont {Lu},
  \citenamefont {Wu},\ and\ \citenamefont {L{\"u}}}]{Lu:2025ntu}%
  \BibitemOpen
  \bibfield  {author} {\bibinfo {author} {\bibfnamefont {K.-P.}\ \bibnamefont
  {Lu}}, \bibinfo {author} {\bibfnamefont {P.-Y.}\ \bibnamefont {Wu}},\ and\
  \bibinfo {author} {\bibfnamefont {H.}~\bibnamefont {L{\"u}}},\ }\bibfield
  {title} {\bibinfo {title} {{Weak Cosmic Censorship Conjecture in Gedanken
  Experiments at All Orders}},\ }\href
  {https://doi.org/10.1103/PhysRevD.111.104036} {\bibfield  {journal} {\bibinfo
   {journal} {Phys. Rev. D}\ }\textbf {\bibinfo {volume} {111}},\ \bibinfo
  {pages} {104036} (\bibinfo {year} {2025})},\ \Eprint
  {https://arxiv.org/abs/2502.02639} {arXiv:2502.02639 [gr-qc]} \BibitemShut
  {NoStop}%
\bibitem{Misner:1973prb}
C.~W. Misner, K.~S. Thorne, J.~A. Wheeler, {Gravitation}, W. H. Freeman, San
  Francisco, 1973.
\end{thebibliography}
%=====================================
\end{document}